\documentclass[onecolumn,trackchanges]{aastex631}

\submitjournal{ApJ}
\accepted{February 18, 2023}
\shorttitle{X-ray Quasi-periodicities of six AGNs}
\shortauthors{Zhang et al.}
\graphicspath{{./}{figures/}}
\begin{document}
\title{Search for X-ray Quasi-periodicity of six AGNs using Gaussian Process method}
\author[0000-0003-3392-320X]{Haoyang Zhang}
\affiliation{School of Physics and Astronomy, Key Laboratory of Astroparticle Physics of Yunnan Province, Yunnan University, Kunming 650091, China}
\author{Shenbang Yang}
\affiliation{Faculty of Science,Kunming University of Science and Technology, Kunming 650500, China}
\author[0000-0001-7908-4996]{Benzhong Dai}
\correspondingauthor{Benzhong Dai}{\email{bzhdai@ynu.edu.cn}}
\affiliation{School of Physics and Astronomy, Key Laboratory of Astroparticle Physics of Yunnan Province, Yunnan University, Kunming 650091,
	China}

\begin{abstract}
The quasi-periodic oscillations (QPOs) found in active galactic nuclei (AGNs) are a very interesting observational phenomenon implying an unknown physical mechanism around supermassive black holes. Several AGNs have been found to have QPO phenomena in the X-ray energy band. Long-duration X-ray observations were collected and reduced for six AGNs with a suspected QPO. The Gaussian process (GP) model \textit{celerite} was used to fit the light curves and to search for the quasi-periodicity behavior. The power spectral density and parameter posterior distributions of each light curve were calculated with the optimal model. Of the six AGNs, only RE J1034+396 was found to have a QPO of about 3600~$s$. The other five sources do not show QPO modulation behavior. We propose that a hot spot on the accretion disk is a possible physical mechanism resulting in this quasi-periodic behavior of AGNs.
\end{abstract}


\section{Introduction} \label{sec:intro}

Quasi-periodic oscillations (QPOs), as a special astrophysical phenomenon seen in all the electromagnetic wavebands, has gradually become a popular topic in recent decades. In X-rays, a QPO appears in accretion systems with a compact central core, such as white dwarfs \citep{1981ApJ...245..618P}, neutron stars \citep{2021MNRAS.500...34T,2022MNRAS.512.4792R}, and stellar mass black holes \citep{1997txra.symp..750R}. The QPO frequencies in these sources generally range from mHz to $10^{2}$ Hz. These sources are collectively known as X-ray binaries with an upper mass limit that is usually on the order of several stellar masses ($5-15M_\odot$). Later, with the development of X-ray telescopes, QPOs were gradually discovered in persistent extragalactic active galactic nuclei (AGNs) \citep{2008Natur.455..369G} which are powered by accretion onto a supermassive black hole (SMBH, $10^6-10^{10}M_\odot$). The frequency of QPOs typically detected in AGNs is extremely low, reaching $\sim10^{-4}$ Hz \citep{2017ApJ...849....9Z,2018ApJ...853..193Z,2020MNRAS.495.3538J}. Our focus is on the X-ray QPOs of AGNs.

These quasi periodic signals are represented by broad peak structures in power spectrum. An inherently periodic signal with a time-dependent period can produce broad peaks in a power spectrum, and a stochastic process with an inherently broadened power spectrum also can produce quasi-periodic behavior, such as a higher-order autoregressive process  \citep{2020ApJS..250....1T}. In addition to the above, we are more concerned that this quasi-periodic behavior may be due to physical processes. In order to explain the physical mechanism of QPOs in AGNs, many models have been put forward, including the most likely orbital resonances models \citep{2001A&A...374L..19A,2003PASJ...55..467A,2008A&A...486....1H}. These models invoked a coupling state consisting of orbital, vertical and radial epicyclic frequencies of particles located on the accretion disk. In the coupled case, if the frequency ratio of the two components is 3:2, a broad peak structure similar to a quasi-periodic signal will be generated in the power spectrum. These models were first applied to the QPOs of X-ray binary, \citet{2015ApJ...798L...5Z} then extended this idea to explain the QPO phenomenon of AGNs. It was suggested that the quasi-periodic phenomena of SMBHs may have precisely the same physical mechanism as the QPOs produced by stellar-mass black holes. The relationship between the periodic frequency and the mass of the central black hole, called the universal scaling relation \citep{2004A&A...425.1075A,2006ARA&A..44...49R,2015ApJ...798L...5Z}, was obtained, which means that we can infer the mass of the black holes from AGN periodicity by the established spin of the black hole. In addition to the orbital resonance model, seismic models, such as pressure ($p$) mode and gravity ($g$) mode, suggest that acoustic pressure and gravity are the restoring forces generated by periodic oscillation in the accretion disk \citep{1987PASJ...39..457O,1997ApJ...476..589P,2003MNRAS.344L..37R}. Astronomers have also have proposed some models with special geometries, such as the hot spot model. In this scenario, by assuming that there are one or more hot spots at a certain radius in the accretion disk surrounding the SMBH, these hot spots move along geodesics near the black hole, generating quasi-periodic radiation \citep{2003APS..APR.P9012S,2004AIPC..714...40S}. Some phenomenological models have also been proposed. For example, a simple dynamical system that exhibits transient chaos called A Dripping Handrail accretion model \citep{1993ApJ...411L..91S,1996ApJ...468..617Y}. In this model, a QPO is generated when accreted material at the edge of the inner disk falls into the accretion disk by an unspecified instability. Similar to A Dripping Handrail, QPO behavior with a certain frequency can also be generated by assuming that the accretion disk spontaneously evolves to A self-organized critical state after material at the inner edge of the disk falls into the center \citep{1994ApJ...435L.125M}.
These models also appear to clarify the observations reported by \citet{2001ApJ...559L..25W,2017ApJ...849....9Z,2018ApJ...853..193Z}. However, \citet{2021ApJ...906...92S} argued that the orbital resonance model is not a perfect fit for the QPO phenomenon of AGNs. Other models have also been met with some skepticism.

Much effort has been dedicated to search for the timescale of QPOs in different energy bands \citep{2008Natur.455..369G,2013MNRAS.436L.114K,2014RAA....14..933Z,2021ApJ...919...58Z}. \citep{2001ApJ...562L.121B} found that Mrk~766 has a lower significant QPO in X-rays. \citet{2012AA...544A..80G} analyzed the X-ray variability data of 104 AGNs and concluded that there was almost no reliable QPO signal. \citet{2016MNRAS.461.3145V} pointed out that QPOs may be false signals and some stochastic process may instead generate periodic signals with 2-5 cycles. For sources reported to have a QPO, \citet{2019MNRAS.482.1270C} and \citet{2020A&A...634A.120A} analyzed $\gamma$-ray data from the Fermi Large Area Telescope (Fermi-LAT), but did not find any QPO phenomenon. Although, QPO timescales are different in $\gamma$-rays and X-rays, the data were all analyzed using traditional Fourier techniques, such as Lomb-Scargle periodogram (LSP; \citet{1976Ap&SS..39..447L}; \citet{1982ApJ...263..835S}) and weighted wavelet Z-transform (WWZ; \citet{1996AJ....112.1709F}). Astronomical observations are time-limited for the monitoring of a single source, which means that taking the Fourier transform of a finite-length time series will produce the red noise leakage \citep{2002MNRAS.332..231U,2003MNRAS.345.1271V}.

Another approach to account for irregular sampling in stochastic light curves is to fit the light curve in the time domain assuming that it is a realization of a Gaussian process (GP) \citep{2009ApJ...698..895K,2010ApJ...708..927K}. This approach is fundamentally different from traditional methods based on Fourier transforms, in that one calculates the likelihood function of the model to fit the original light curve directly rather than applying a Fourier transform the original data. Compared with Fourier technique, the method of fitting light curve directly can deal with the problem of unequal time interval sampling in astronomical time series. By fitting the light curve and determining the model parameters, the stochastic process can effectively avoid the red noise leakage. This is known as a first-order continuous-time autoregressive process (CAR(1)), introduced by \citet{2009ApJ...698..895K} to fit astronomical time series. The general class of continuous-time autoregressive moving average (CARMA, \citet{2014ApJ...788...33K}) developed from CAR(1), is a GP model suitable for astronomical time series analysis. It has been applied to analyze the astronomical time domain observations \citep{2018ApJ...863..175G,2019ApJ...885...12R}.

With the dramatic increase in astronomical data, CARMA is limited by computational cost and scaling. \citet{2017AJ....154..220F} developed a novel method, \textit{celerite}\footnote{\url{https://celerite.readthedocs.io/en/stable/}}, which computes a class of GP models directly and exactly that scales linearly with the number of data points for one-dimensional datasets. The advantage of \textit{celerite} is that it deals with discrete data sampled at unequal intervals when finding variations by fitting light curves. Compared to the CARMA method, \textit{celerite} is much faster with an average difference of about one order of magnitude. In addition, \textit{celerite} is more scalable than CARMA, users can choose or build their own covariance functions models for different analysis purposes. This advantage will help us to explore the real physical mechanism of AGN \citep{2021Sci...373..789B,2022ApJ...930..157Z}. In our previous work, this method was used to fit the $\gamma$-ray light curves in AGN, and it demonstrated better reliability than traditional LSP \& WWZ methods \citep{2021ApJ...907..105Y}.

In this paper, the \textit{celerite} GP model is employed to reanalyze the X-ray light curves of six AGNs suspected to produce QPOs. Section \ref{sec:datared} describes the extraction of the light curves of six AGNs. In Section \ref{sec:dataanalysis}, the \textit{celerite} tool is introduced. The analyzed results are given in Section \ref{sec:results}. Section \ref{sec:disandcon} contains a brief discussion of the possible physical mechanism of the QPO.

\section{Data Reduction} \label{sec:datared}
The present work focuses on the search for the X-ray QPOs of AGNs using a novel and different approach. The properties of the QPOs and the properties of six AGNs with either suspected or confirmed QPOs are shown in Table \ref{tab:data}. The six objects belong to a class of AGN known as narrow-Line Seyfert 1 (NLS1) galaxies, characterized by broad Balmer lines with low velocity widths, significant Fe~II emission, and a very soft X-ray spectrum \citep{2008ApJ...680..926K}. Such sources are the good candidates for searching for QPOs, since they have much longer X-ray observations, relatively high luminosities, and high-amplitude X-ray variability.

\begin{deluxetable*}{cccccc}[t]
	\tablenum{1}
	\tablecaption{Observations of X-ray QPOs around six AGN \label{tab:data}}
	\tablewidth{0pt}
	\tablehead{
		\colhead{Source Name} & \colhead{Mass} & \colhead{BH Spin} & \colhead{QPO } & \colhead{QPO significanc} &\colhead{$M_{BH}$ \& Spin \& QPO References}\\
		\colhead{} & \colhead{$M_{\sun}$} & \colhead{$a$} & \colhead{$s$}
	}
	
	\startdata
	ESO 113-G010 & $\sim7\times10^6$ & 0.998 & 8065,14724 & 7.3$\sigma$, 4.8$\sigma$ & 1,10 \\
	1H 0707-495 & $\sim2.4\times10^6$ & \textgreater0.976 & 3846,8240 & \textgreater$3\sigma$, $5\sigma$ & 2,7,11,12 \\
	RE J1034+396 & $\sim1\times10^7$ & 0.998 & 3703.7 & $9\sigma$ (at obs 2018)&3,13  \\
	Mrk 766 & $\sim6.6\times10^6$ & \textgreater0.92 & 4200,6452 &...,$5\sigma$ & 4,8,14,15 \\
	MCG-06-30-15 & $\sim1.5\times10^6$ & \textgreater0.917 & 3663 & $3\sigma$ & 5,9,16 \\
	MS 2254.9-3712 & $\sim3.9\times10^6$ & ... & 6666.7 & $3.3\sigma$& 6,17\\
	\enddata
	\tablecomments{
		References: (1) \citet{2013ApJ...764L...9C}, (2) \citet{2005ApJ...618L..83Z}, (3) \citet{2016AA...594A.102C}, (4) \citet{2015PASP..127...67B}, (5) \citet{2016ApJ...830..136B}, (6) \citet{2010ApJS..187...64G}, (7) \citet{2010MNRAS.401.2419Z}, (8) \citet{2018MNRAS.480.3689B}, (9) \citet{2007PASJ...59S.315M}, (10) \citet{2020AcASn..61....2Z}, (11) \citet{2016ApJ...819L..19P}, (12) \citet{2018ApJ...853..193Z}, (13) \citet{2020MNRAS.495.3538J}, (14) \citet{2001AA...365L.146B}, (15) \citet{2017ApJ...849....9Z}, (16) \citet{2018AA...616L...6G}, (17) \citet{2015MNRAS.449..467A}.}
\end{deluxetable*}

Long-duration X-ray data were used to analyze the QPOs. Details of the \textit{XMM-Newton} X-ray observations of the six objects are shown in Table \ref{tab:datailsobs}, which lists all the long-duration observations. All data were analyzed for sources with multiple, long-duration observations.

The \emph{XMM-Newton} satellite has two CCDs, PN and MOS cameras. The PN detector has almost no photon loss and more high-energy sensitivity at 0.2-10 KeV. Thus, only X-ray data from the European Photon Imaging Camera (EPIC) PN detector were analyzed here \citep{2001A&A...365L..18S,2001A&A...365L..27T}. For the data reduction, process, we used the standard procedure of the \textit{XMM-Newton} Science Analysis System (SAS) v19.0.0\footnote{\url{https://www.cosmos.esa.int/web/xmm-newton/download-and-install-sas}} and latest Valid CCF\footnote{\url{https://www.cosmos.esa.int/web/xmm-newton/current-calibration-files}}, set as data reduction software and calibration files.

\begin{deluxetable*}{cccccccc}[t]
	\tablenum{2}
	\tablecaption{Details of observation for 6 AGNs  \label{tab:datailsobs}}
	\tablewidth{0pt}
	\tablehead{
		\colhead{Source Name} & \colhead{Obs.ID} & \colhead{Mode} & \colhead{Expo.ID} & \colhead{Filter} & \colhead{Start Time} & \colhead{Duration} & \colhead{Pile-up}\\
		\colhead{} & \colhead{(1)} & \colhead{(2)} & \colhead{(3)} & \colhead{(4)} & \colhead{(5)} & \colhead{(6)} & \colhead{(7)}
	}
	
	\startdata
	ESO 113-G010 & 0301890101 & Large Window & S003 & MEDIUM & 2005-11-10 & 101855 & yes\\ \hline
	1H0707-495 & 0506200301 & Large Window & S003 & MEDIUM & 2007-05-14 & 38670 & no\\
	& 0511580401 & Large Window & U002 & MEDIUM & 2008-02-04 & 101839 & no\\
	& 0653510401 & Large Window & S003 & MEDIUM & 2010-09-15 & 125855 & yes\\
	& 0653510501 & Large Window & S003 & MEDIUM & 2010-09-17 & 117045 & yes\\ \hline
	RE J1034+396 & 0506440101 & Large Window & S003 & THIN1 & 2007-05-31 & 91393 & yes\\
	& 0655310101 & Small Window & S003 & THIN1 & 2010-05-09 & 44372 & no\\
	& 0655310201 & Small Window & S003 & THIN1 & 2010-05-11 & 53011 & no\\
	& 0675440101 & Small Window & S001 & THIN1	& 2011-05-27 & 36210 & no\\
	& 0675440201 & Small Window & S001 & THIN1 & 2011-05-31 & 29420 & no\\
	& 0824030101 & Small Window & S003 & THIN1 & 2018-10-30 & 71640 & no\\ \hline
	Mrk 766 & 0096020101 & Small Window & S012 & MEDIUM & 2000-05-20 & 38969 & no\\
	& 0304030601 & Small Window & S003 & MEDIUM & 2005-05-31 & 98471 & no\\ \hline
	MCG-06-30-15 & 0111570201 & Small Window & S003 & MEDIUM & 2000-07-11 & 55002 & no\\ \hline
	MS 2254.9-3712 & 0205390101 & Small Window & S003 & MEDIUM & 2005-05-01 & 70972 & no\\
	\enddata
	\tablecomments{(1) Observation ID, (2) observation mode, (3) exposure ID, (4) used filter, (5) start time of the observation, (6) total duration time, (7) pile-up effect.}
\end{deluxetable*}

In the specific process, we used the \textit{epproc} script to reprocess the Observation Data Files (ODFs) to obtain and to concatenate event lists. In order to select Good Time Intervals, the \textit{tabgtigen} script was used. The background count-rate threshold was ``RATE\textless=0.4", used to determine where the light curve was low and steady. Next, we used \textit{evselect} to filter out the flaring particle background using PN event lists to produces clean photon files. The ranges of the source center area for the Small and Large Window observation mode were generally 40 and 50 arcsec, respectively. In Large Window mode, the background photon-counts region was an annulus, ranging between 50 and 100 arcsec. In Small Window mode, the region is circular, with a radius of 40 arcsec far away from the source on the CCD chip. Before extracting the final photon counts file, the pile-up effect must be checked with the \textit{epatplot} script. Finally, the \textit{evselect} and \textit{epiclccorr} script were used to extract the light curve by subtracting the background. The results in Table \ref{tab:data} are basically obtained by using 100~s or 200~s bins light curves. In order to avoid accidental errors caused by minimum time interval selection, time bins were selected as $100$ s and $200$ s for extracting the light curves.

\section{Data Analysis method} \label{sec:dataanalysis}
\subsection{Gaussian process in celerite} \label{sec:GP}
The basic principle of GP is to calculate the likelihood function of the model, so as to determine the parameters of the model and complete the fitting of the curve \citep{2006GPML}.
This subsection mainly describes \textit{celerite} as a tool to implement GP, how to choose the best model, and judge the goodness of fit.

For a one-dimensional time series in astronomy $y=f(t)+\sigma_{err}$, where $\sigma_{err}$ is the measurement error. \textit{celerite} follows the basic assumption of GP. A GP model consists of a mean function $\mu_{\theta}(t)$, and a covariance function $k_{\alpha}(t_{n},t_{m})$ with the parameters $\theta$ and $\alpha$, respectively. Thus, the log-likelihood function $\mathcal{L}$($\theta$,$\alpha$), of a GP model is:
\begin{equation}
	\ln\mathcal{L}(\theta,\alpha)= -\frac{1}{2}r_{\theta}^{T}K_{\alpha}^{-1}r_{\theta}-\frac{1}{2}\ln \det K_{\alpha}-\frac{N}{2}\ln(2\pi)  \label{equ:likelihood}
\end{equation}
where $r_{\theta}=y-\mu_{\theta}(t)$, $[K_{\alpha}]_{mn}=\sigma_{err}^{2}\delta_{mn}+k_{\alpha}(t_{n},t_{m})$, $\delta_{mn}$ is the Kronecker delta symbol, and $N$ is the number of data points. The estimation of parameters $\theta$ and $\alpha$ can be achieved by the maximum likelihood method and Equation \ref{equ:likelihood}.

In general, $\mu_{\theta}(t)$ is usually set to the mean of the original observed data. Thus the core of GP is to solve the covariance function $k_{\alpha}(t_{n},t_{m})$. Each covariance function represents a different model, so it is important to choose the appropriate covariance function for the physical properties being fit.

The \textit{celerite} tool builds a mixture of exponentials covariance function as:
\begin{equation}
	k_{\alpha}(t_{m},t_{n})=\sigma_{err}^{2}\delta_{mn}+\sum_{j=1}^{J}a_{j} \exp(-c_{j}\tau_{mn}) \label{equ:kernel}
\end{equation}
where $J$ is the number of components in the mixture and $\tau_{mn}\equiv|t_{m}-t_{n}|$.
By exploiting the semiseparable structure of these covariance matrices, the covariance function can be efficiently and quickly calculated using the Cholesky factorization method. The cost of computing the covariance matrix can be reduced from $O(N^{3})$ for the standard GP to $O(NJ^{2})$ in \textit{celerite}, where $N$ is the number of data points \citep{2017AJ....154..220F}.

\begin{figure*}[th]
	\figurenum{1}
	\gridline{\fig{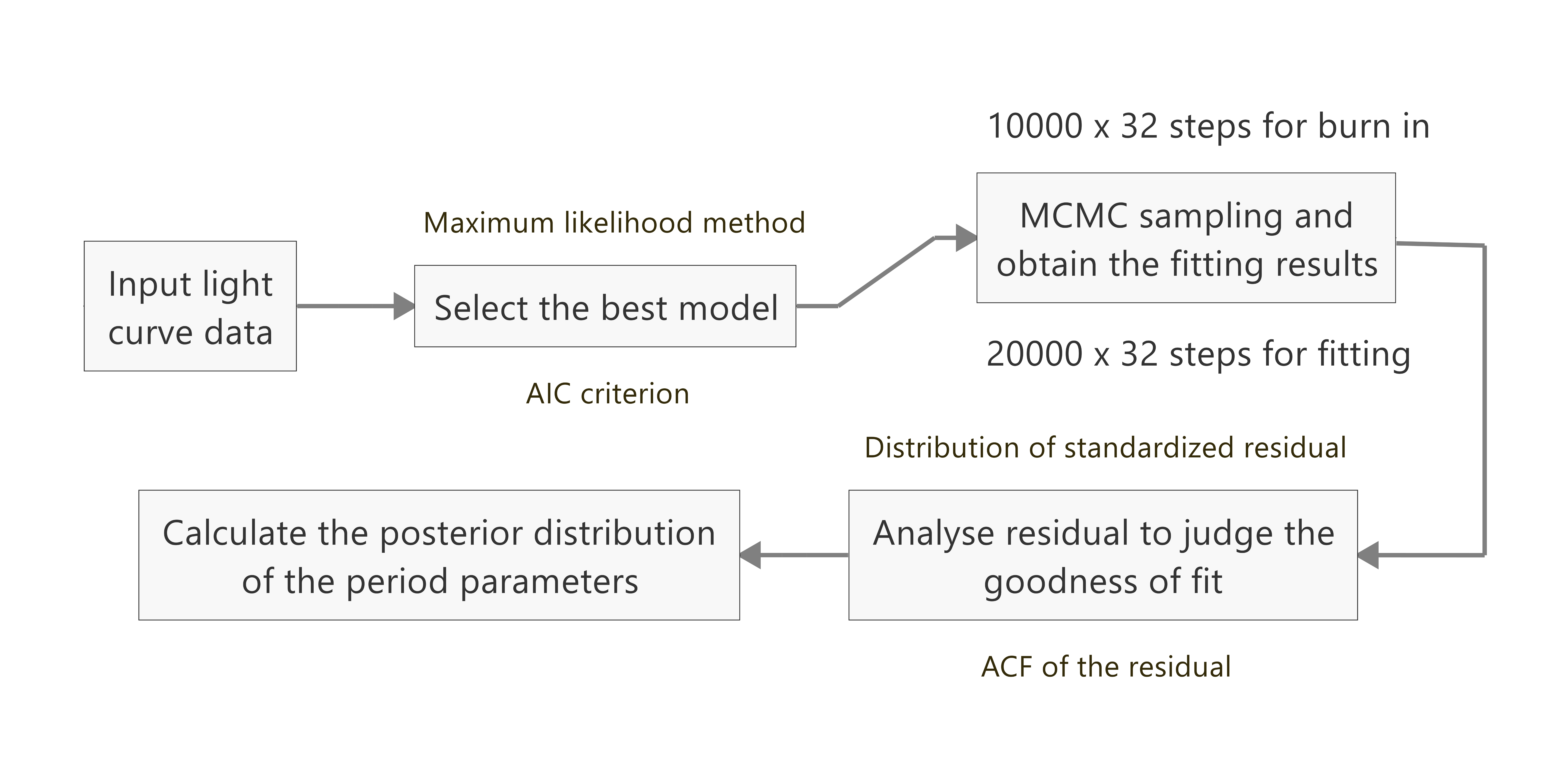}{\textwidth}{}
	}
	\caption{A complete flow chart of the fitting process. \label{fig:GPflow}}
\end{figure*}

In the next subsection, we will introduce a special covariance function with intrinsic periodic properties. If a single model describes a phenomenon that is too simple, we usually use the superposition of multiple models as a new model \citep{2006GPML,2017AJ....154..220F}. As the expressions of these superposition models are unknown, \textit{celerite} can concatenating the parameters and coefficients in the correct order \footnote{\url{https://celerite.readthedocs.io/en/stable/python/kernel/}}. By selecting different numbers of model terms in the superposition, we can construct a variety of different candidate models.

In our fitting process, we provide multiple candidate models. According to the suggestions of \citet{2014ApJ...788...33K} and \citet{2017AJ....154..220F}, for the GP model, we can use an information criterion to screen out the optimal model to fitting. The Akaike Information Criterion (AIC) is used to select the most reasonable model. AIC does not need to calculate the posterior probability or posterior distribution of the parameters, thus this method can quickly and effectively select the best model. AIC is defined as \citep{2014ApJ...788...33K}:
\begin{equation}
	AIC=2k-2log\ \mathcal{L}+\frac{2k(k+1)}{n-k-1}
\end{equation}
where $k$ is the number of models parameters, $n$ is the number
of data points on the light curves, and $\mathcal{L}$ is the maximum likelihood from Equation \ref{equ:likelihood}. To avoid the large uncertainties of the maximum likelihood (i.e. instability of the L-BFGS-B optimizer in the \textit{celerite} mothod), each model is fitted 100 times with different initial parameters. The model with the minimum AIC value is selected as the best-fitting model of the light curve.

Although using an information criterion is an approximate method compared with other model selection methods, in the GP model, the parameter $\alpha$ of the covariance function is set it to Gaussian distribution, which exactly conforms to the assumption of the information criterion.

In Bayesian theory, the prior distributions of model parameters can be multiplied by the likelihood function to obtain the posterior distributions of model parameters. This is done in the final fitting process after selecting the best model. We have embedded the Markov Chain Monte Carlo (MCMC) sampler \texttt{emcee}\footnote{\url{https://emcee.readthedocs.io/en/v2.2.1/}} \citep{2013PASP..125..306F} in \textit{celerite} to obtain the posterior distributions of the model parameters and fit the light curve of each observation. \texttt{emcee} is run using 32 parallel walkers for 10000 steps as burn-in and 20000 steps for MCMC sampling to obtain the best-fitting curve.

Since the astronomical time series contains measurement errors, the distribution of errors is a normal distribution with a mean of zero. If the model is correct, then the errors are modeled as residials and obey the same distribution law with measurement errors. Therefore, \citet{2014ApJ...788...33K} confirmed that for a well-fitted GP model, the standardized residuals and squared standardized residuals should be irregular sequences of white noise, following a Gaussian distribution with $\mu=0$, $\sigma=1$. Thus the Kolmogorov-Smirnov (KS) test is then used to judge the correlation between the residuals of the best fit and a Gaussian distribution. In the KS test, a value of P\textgreater0.05 indicates a good fit. The quality of the fitting results determines the reliability of the results.

In order to verify the reliability of the fitting results by several different methods, we also calculate the auto-correlation function (ACF) of the standardized residuals, squared standardized residuals, and the 95\% confidence interval of the corresponding white noise for comparison. If most of the ACF points are outside the 95\% confidence interval of the white noise, it means that the fitting results may be unreliable.

We have also presented a graphical fitting flow chart to describe our fitting process (Figure \ref{fig:GPflow}). In the process of analyzing each observation, we strictly followed the process described above in the analyses, and ensure that the interface between \texttt{emcee} and \textit{celerite} is correctly configured and invoked.

\subsection{A celerite model}
The power spectral density (PSD) is the variability amplitude per frequency. It describes the variability power contained within a frequency interval. The variability of quasi-periodic and stochastic light curves is often characterized through the PSD \citep{2014ApJ...788...33K}. After fitting the light curve with \textit{celerite}, we can determine the parameter distribution of the covariance function (model), where the covariance function and PSD are usually Fourier transform pairs. The GP model \textit{celerite} is used to fit the X-ray light curve and to calculate the PSD.

The variability of oscillations can be excited by noisy physical processes and grow strongly at a characteristic timescale, but is also damped owing to dissipation in the system. Equation \ref{equ:kernel} presents a formal expression of the model. In the specific case, \textit{celerite} considers the dynamics of a stochastically driven damped simple harmonic oscillator (SHO, \citet{2017AJ....154..220F}). The differential equation of the SHO is expressed as:

\begin{equation}
	[{\frac{d^2}{dt^2}}+{\frac{\omega_0}{Q}\frac{d}{dt}}+{\omega_0}^2]\ y(t)=\epsilon(t)
\end{equation}
where $\omega_0$ is the frequency of the undamped oscillator, $Q$
is the quality factor of the oscillator, and $\epsilon(t)$ is assumed
to be white noise. The PSD of this process is:

\begin{equation}
	S(\omega)=\sqrt{\frac{2}{\pi}}{\frac{S_0{\omega_0}^4}{(\omega^2-{\omega_0}^2)^2+{\omega_0}^2\omega^2/Q^2}}  \label{equ:psd}
\end{equation}
where $S_0$ is proportional to the power at $\omega={\omega}_0$ and  $S(\omega_0)=\sqrt{2/\pi}S_0Q^2$. Therefore, $\omega_0$ is the peak position in the whole PSD, which exactly corresponds to the frequency of QPO. The PSD peak frequency indicates a suspect QPO and the $Q$ value at the peak location should be greater than 0.5 \citep{2017AJ....154..220F}. This model contains inherent periodicity ($\omega_0$) and is a rationalization model to describe QPO behavior.

According to Section \ref{sec:GP}, a single SHO model may not be able to fit a complex astronomical time series well.  \citet{2017AJ....154..220F} indicates that complex models can be expressed by superimposing simple SHO models. In order to find the broad peak structure in the PSD better, we adopt the superposition of damped random walk\footnote{In some papers, this model is referred to as the Uhlenbeck-Ornstein process} (DRW) model and SHO model to fit the light curves \citep{2021ApJ...907..105Y}. DRW is used to represent red noise components of the PSD \citep{2018ApJ...859L..12L}. Superimposing this model can make the fitted PSD closer to the PSD of the real data. Two types of model, m$\times$SHO and DRW+m$\times$SHO, are applied to fit the observational light curves. Here, m is the number of terms in the superposition. In general, a model with terms $\leq4$ is sufficient to fit the data. We note that more terms may bring more parameters, which will complicate the model and is not conducive to the fitting process. In total, for each light curve, eight candidate models are used in the fitting process.

\section{Results} \label{sec:results}
We used \textit{celerite} to fit the observations in Table \ref{tab:datailsobs}. Among the six AGNs previously reported to have a suspicious QPO, in only two observations (ObsID:0506440101 \& ObsID:0824030101) of RE J1034+396 did we find obvious peak structure in the corresponding PSDs. We have plotted the fitting results of these two observations and the PSDs separately, and the analysis results of the other sources are in the APPENDIX \ref{sec:detailsandpsd}. Figure \ref{fig:fit1} and Figure \ref{fig:fit2} present the fitting results of ObsID:0506440101 and ObsID:0824030101, respectively. The models used to fit of each observation and the fitting results of all sources are shown in Table~\ref{tab:reduction}. In Table~\ref{tab:reduction}, the first column is the source name, the second column is the observation ID, columns 3, 4 are the best-fit models, the KS test values for the light curve with $100~s$ time bin, respectively. Columns 5 and 6 shwo the same but for $200~s$ time bin.

In Figure \ref{fig:fit1}, panel (a) and panel (c) show the best-fitting curves and the $1\sigma$ confidence intervals for observations with $100s$ and $200s$ bins by the MCMC sampler, respectively. The following is a judgment of the goodness of fit. The standardized residuals of the best fits are shown in panel (b) and panel (d) of Figure \ref{fig:fit1}.  Histograms of the standardized residual distribution are approximated as a Gaussian distribution with $\mu=0$ and $\sigma=1$. KS test were then used to judge the correlations between the residuals of the best fit and a normal distribution. We also calculated the ACFs of the standardized residual and squared standardized residuals. It was found that the ACF results were mostly in the 95\% white noise confidence interval, indicating that \textit{celerite} fitted the light curves well. In Figure \ref{fig:fit1}, the panel (e) and panel (f) present the ACFs of the standardized residuals and squared standardized residuals for the $100~s$ time bin data, while panel (g) and panel (h) show the same for the $200~s$ time bin data. The grey region is the 95\% white noise confidence interval. The fitting results of the other sources are presented according to the same rules as in Figure \ref{fig:fit1}.

The PSDs were calculated after fitting the light curves using Equatiaon \ref{equ:psd}. The measurement noise level is estimated by:
\begin{equation}
	P_{noise}=2\times\triangle t\times \overline{\sigma_{err}^{2}}
\end{equation}
where $\triangle t$ is the average time interval between two data points or the selected time bin of the light curve and $\sigma_{err}$ is the measurement error.

\begin{figure*}[h]
	\figurenum{4}
	\gridline{\fig{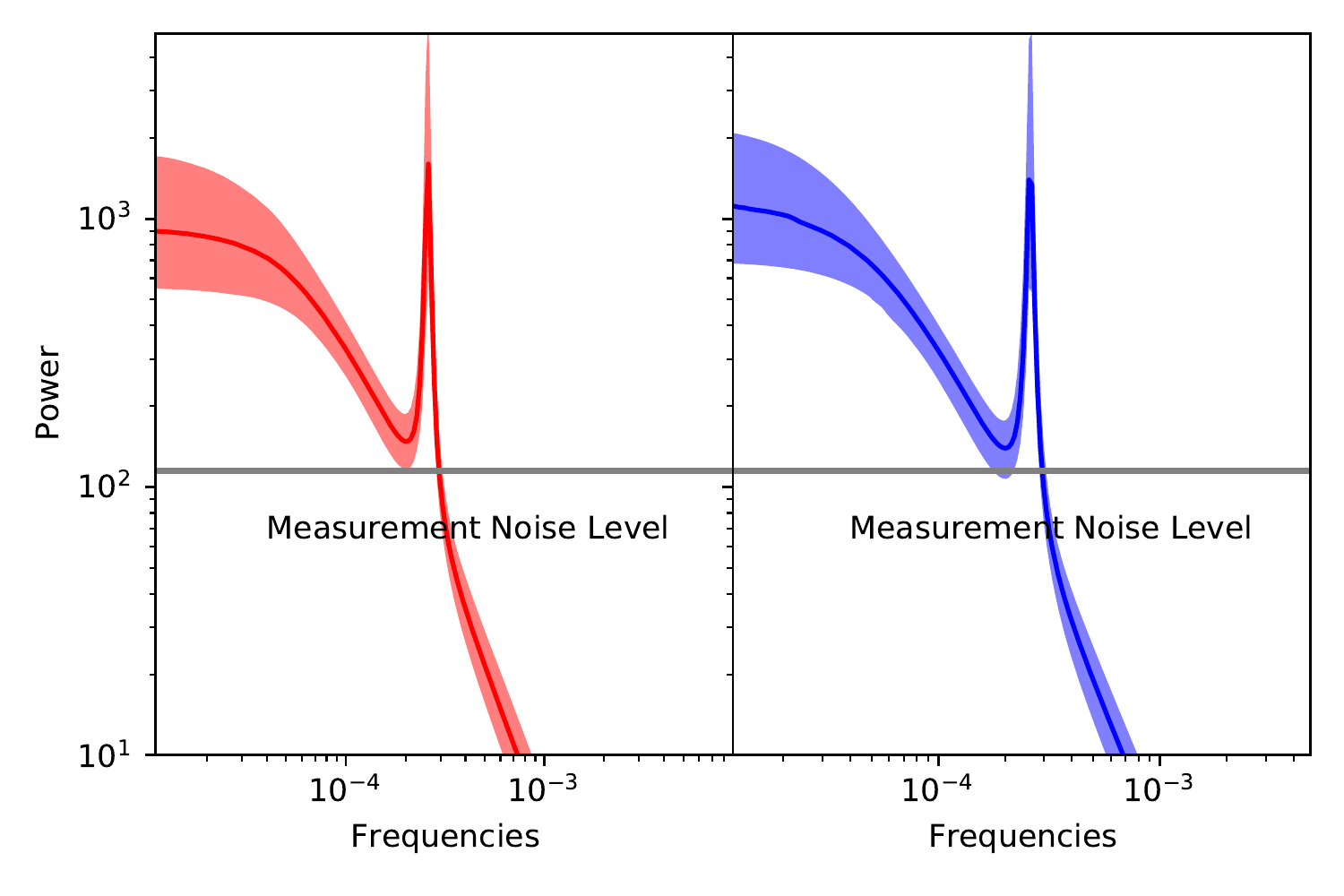}{0.7\textwidth}{}
	}
	\caption{The PSD of ObsID:0506440101 for RE~J1034+396. The red line and its 1$\sigma$ standard deviation represent the results for the 100s time bin data, while the blue one is for the 200s time bin data. The gray line is the measurement noise level. The peaks imply the QPO signals. \label{fig:psd1}}
\end{figure*}

The PSD of the fitting result for RE J1034+396 in two observations of ObsID:0506440101 and ObsID:
0824030101 are shown in Figure \ref{fig:psd1} and Figure \ref{fig:psd2}, respectively. The PSDs for all sources and all observations are shown in APPENDIX \ref{sec:detailsandpsd}, except for the above two observations. In general, the shapes of the PSD distributions can be approximated as a bending or broken power-law spectrum. The suspected QPO signal is located at the peak of PSD. According to our analysis, only RE J1034+396 was found to have suspected QPOs. The median value of the PSD, which always indicates the QPO signal, obviously exceeds the measurement noise level.

\begin{figure*}[t]
	\figurenum{6}
	\gridline{\fig{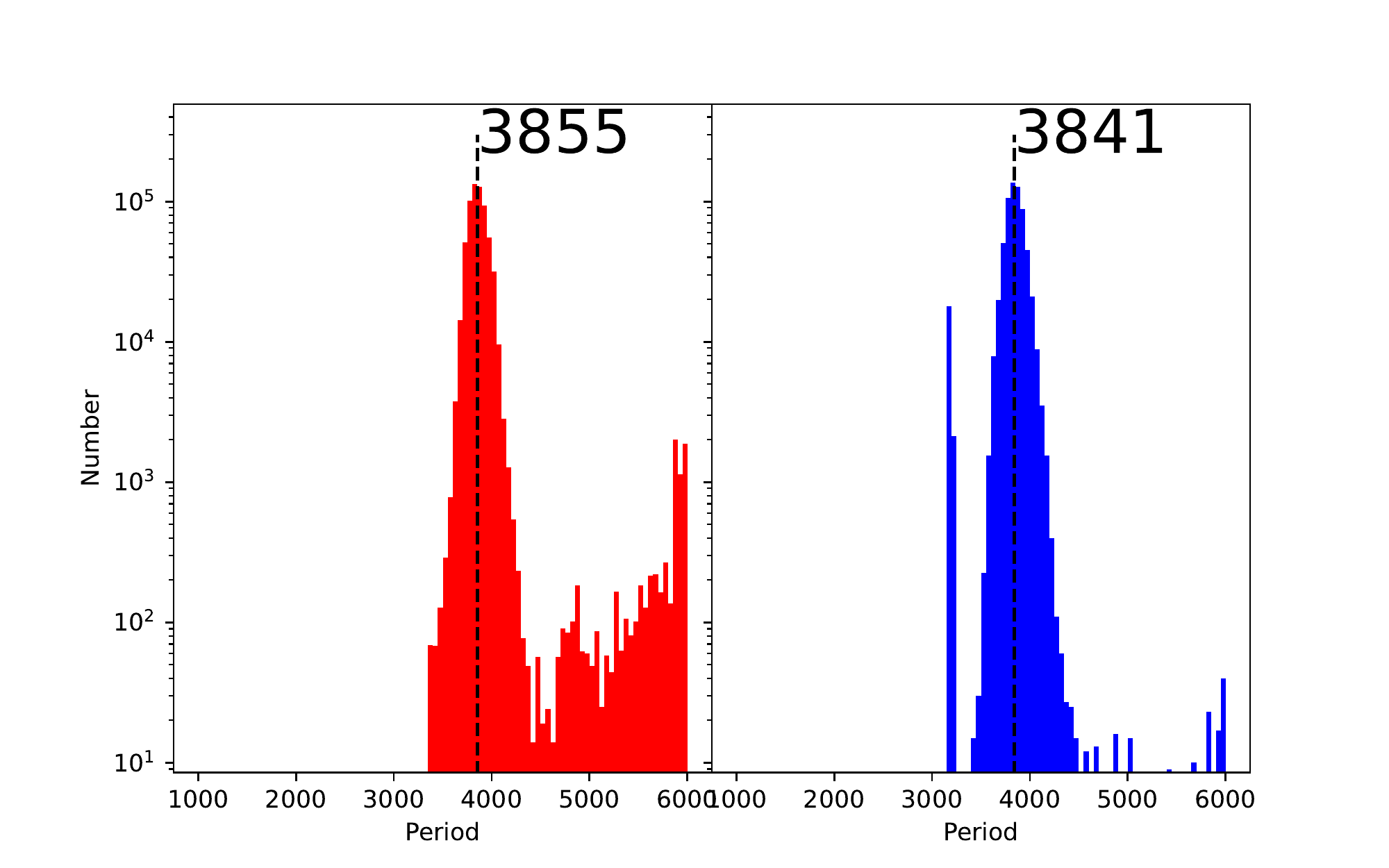}{0.5\textwidth}{}
		\fig{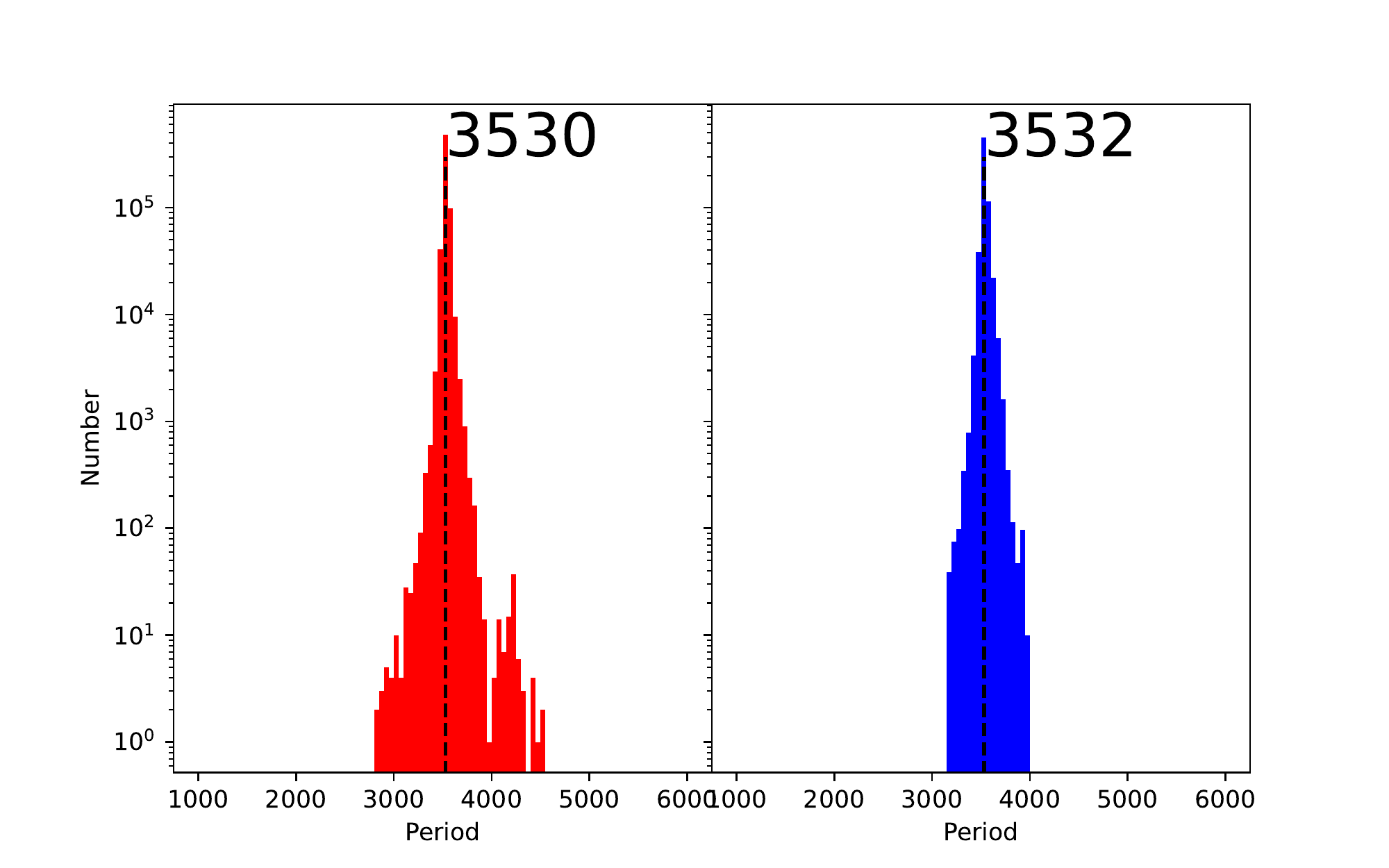}{0.5\textwidth}{}
	}
	\caption{The posterior distributions of the period, $T_{period}$, of ObsID:0506440101 (left) and ObsID:0824030101 (right) for RE~J1034+396. The red one represents 100s binned data and blue one represents 200s binned data. The black dotted line is the median value of the distribution which corresponds to the QPO period. \label{fig:distr}}
\end{figure*}

To evaluate the significance of the possible OPQ signal, the posterior distribution of the parameter $\omega_0$ was calculated in accordance with basic GP theory. In order to display the results of the period more intuitively, we use $T_{period}={2\pi}/{\omega_0}$ to derive the results of the period. The posterior distribution of $32\times20000$ MCMC results is shown in Figure \ref{fig:distr}. The high significance would be expected to follow a Gaussian distribution. The details of the suspected QPOs of RE~J1034+396 are shown in Table \ref{tab:result}.

Table \ref{tab:result} shows two high-significance QPOs for RE~J1034+396. One is in ObsID:0506440101 and the another in ObsID:0824030101. The QPO with larger error is $~$3800 s, and it is $~$3500 s with smaller errors. The parameter $Q$ is taken from the median sampled by MCMC sampler. The value of the quality factor of the oscillator $Q$ reflects the stability of the period value. Larger $Q$ values indicate more stable QPO signals.

\begin{figure*}[th]
	\figurenum{2}
	\gridline{\fig{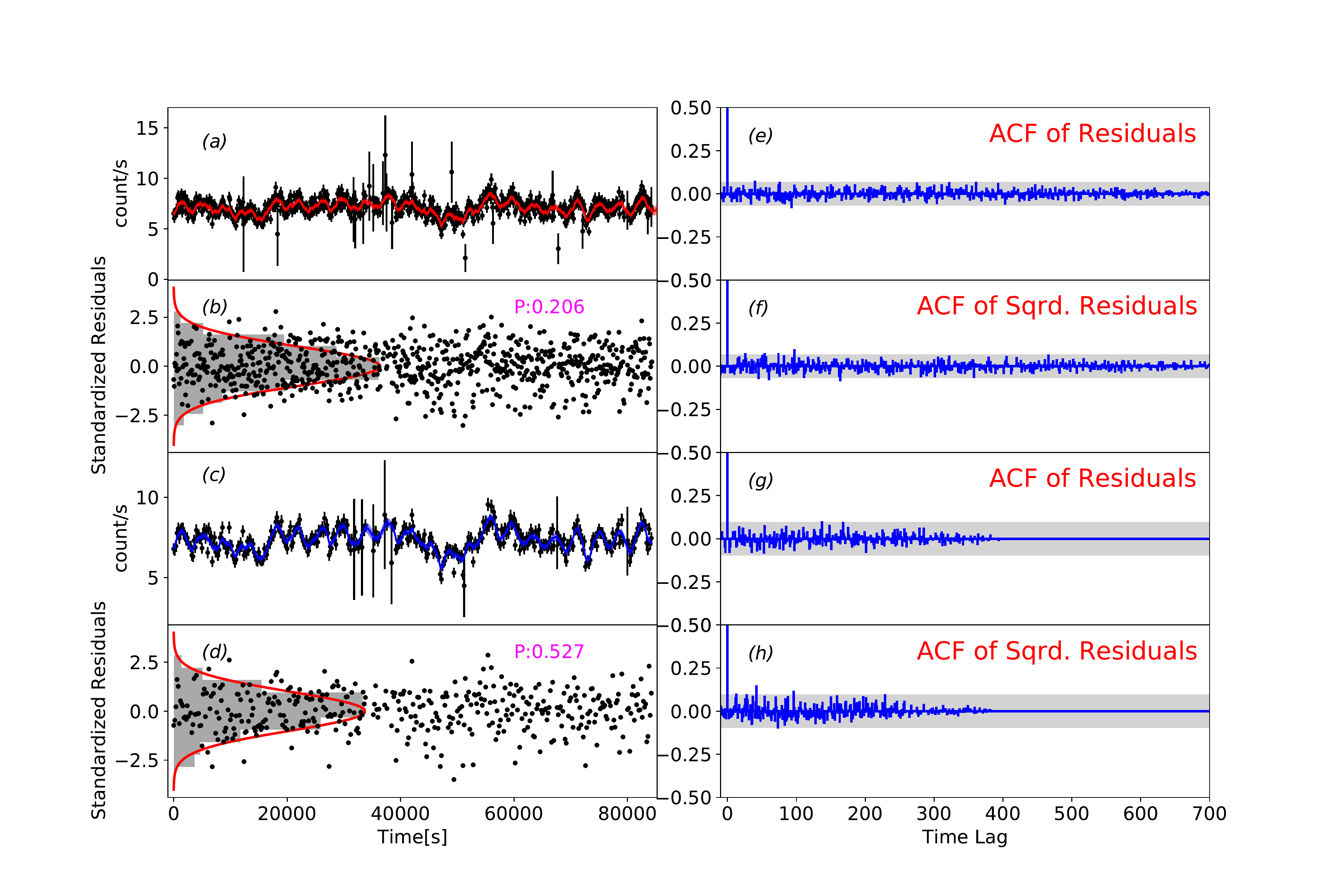}{\textwidth}{}
	}
	\caption{Light curve fitting of ObsID:0506440101 for RE~J1034+396. Panel (a) and (c) show the light curve and the best fitting with a 1$\sigma$ deviation for 100s and 200s time bin data, respectively. Panel (b) and (d) are the distributions of the standardized residuals. A Gaussian distribution with $\mu=0$ and $\sigma=1$ is shown (red line). The P values in panel (b) and panel (d) are the KS test values. Panels (e) and (f) are the ACFs of residuals for the 100s and 200s time bin data, while Panels (g) and (h) are the ACFs of the squared residuals for 100s and 200s time bin data. The gray area is the 95\% confidence interval of white noise. \label{fig:fit1}}
\end{figure*}

\begin{figure*}[th]
	\figurenum{3}
	\gridline{\fig{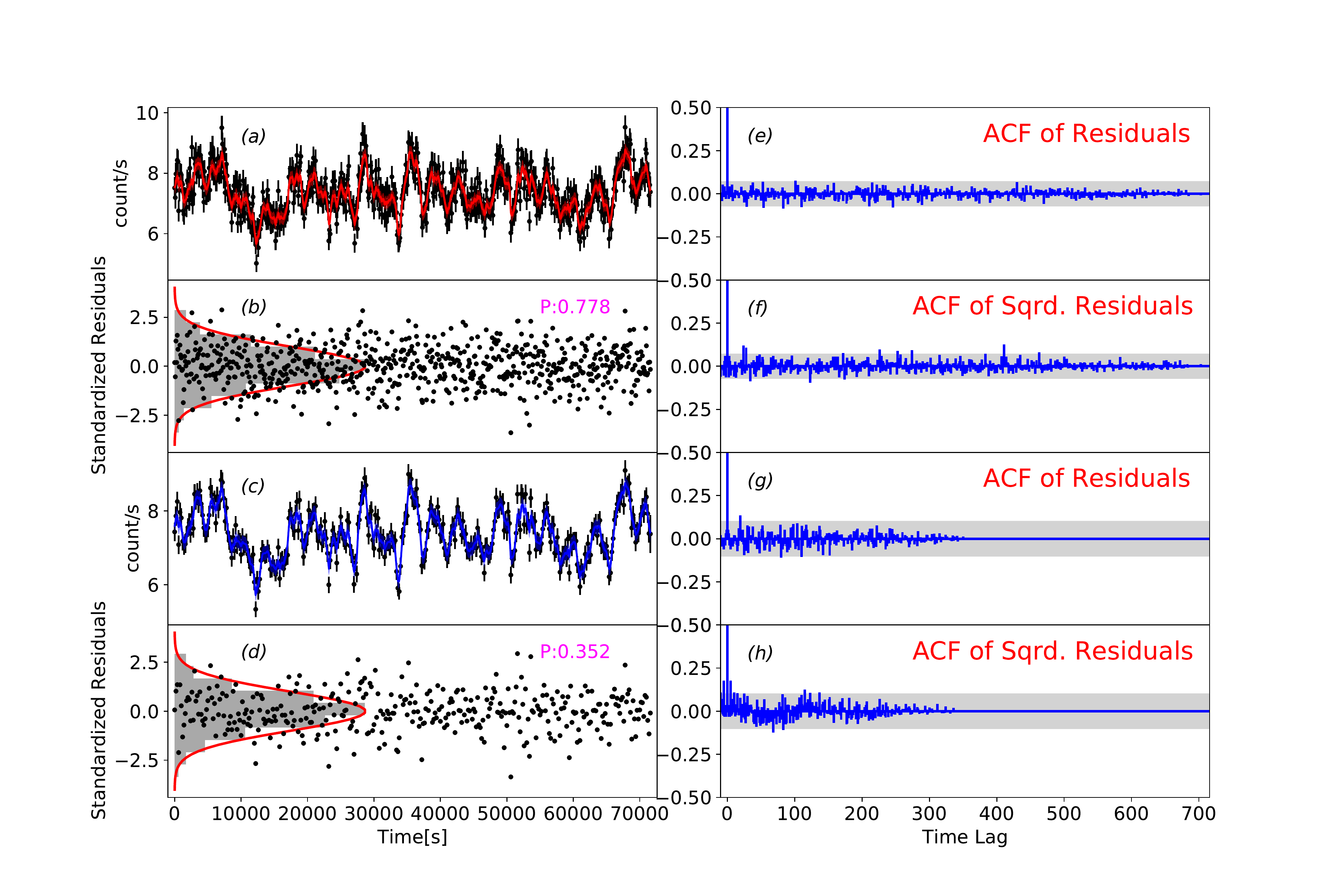}{\textwidth}{}
	}
	\caption{Same as Figure \ref{fig:fit1}, but for ObsID:0824030101 of RE~J1034+396. \label{fig:fit2}}
\end{figure*}

\begin{deluxetable*}{cccccc}[b]
	\tablenum{3}
	\tablecaption{The data analysis results for all sources and all selected observations. \label{tab:reduction}}
	\tablewidth{0pt}
	\tablehead{
		\colhead{Source Name} & \colhead{ObsID} & \colhead{\textit{celerite} model(100 s/bin)} & \colhead{p-value(100 s/bin)} & \colhead{\textit{celerite} model(200 s/bin)} & \colhead{p-value(200 s/bin)} \\
	}
	
	\startdata
	ESO 113-G010 & 0301890101 & 4$\times$SHO & 0.090 & DRW+4$\times$SHO & 0.210 \\ \hline
	1H0707-495 & 0506200301 & DRW+SHO & 0.482 & 2$\times$SHO & 0.197 \\
	& 0511580401 & 2$\times$SHO & 0.503 & 2$\times$SHO & 0.255 \\
	& 0653510401 & 2$\times$SHO & 0.083 & 2$\times$SHO & 0.664 \\
	& 0653510501 & 3$\times$SHO & 0.036 & 2$\times$SHO & 0.140 \\ \hline
	RE J1034+396 & 0506440101 & 4$\times$SHO & 0.310 & 4$\times$SHO & 0.658 \\
	& 0655310101 & DRW+2$\times$SHO & 0.192 & 2$\times$SHO & 0.792 \\
	& 0655310201 & DRW+SHO & 0.260 & DRW+SHO & 0.549 \\
	& 0675440101 & DRW+SHO & 0.821 & DRW+2$\times$SHO & 0.810 \\
	& 0675440201 & 2$\times$SHO & 0.282 & DRW+SHO & 0.969 \\
	& 0824030101 & DRW+2$\times$SHO & 0.778 & DRW+3$\times$SHO & 0.352 \\ \hline
	Mrk 766 & 0096020101 & 3$\times$SHO & 0.262 & 2$\times$SHO & 0.939 \\
	& 0304030601 & 2$\times$SHO & 0.421 & 3$\times$SHO & 0.429 \\ \hline
	MCG-06-30-15 & 0111570201 & 2$\times$SHO & 0.572 & 2$\times$SHO & 0.683 \\ \hline
	MS 2254.9-3712 & 0205390101 & DRW+SHO & 0.414 & DRW+SHO & 0.654 \\
	\enddata
	\tablecomments{The p-value is the KS test of standardized residuals of fitting results, that should be $\geq$\,0.05. }
\end{deluxetable*}

\begin{figure*}[h]
	\figurenum{5}
	\gridline{\fig{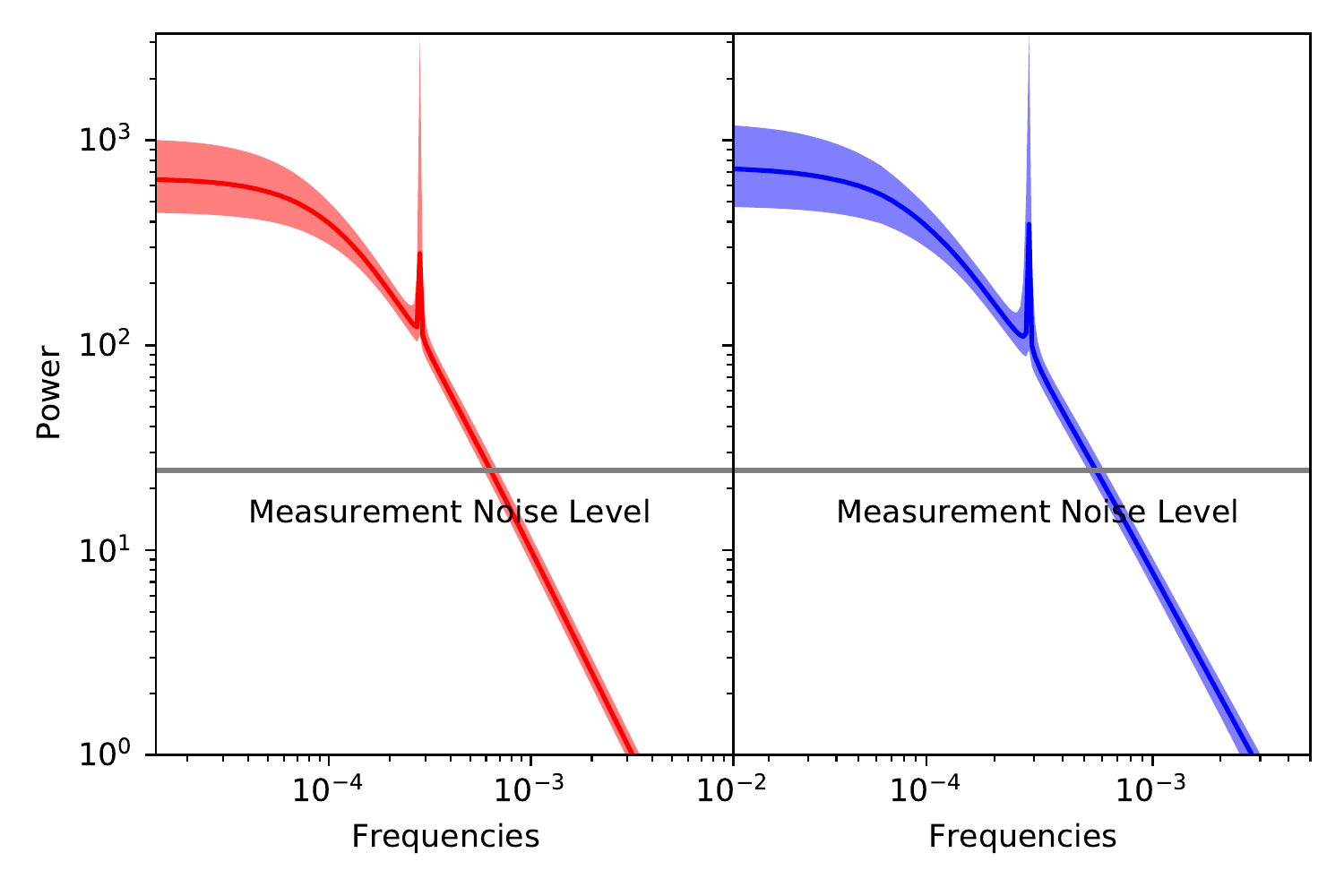}{0.7\textwidth}{}
	}
	\caption{Same as Figure \ref{fig:psd1}, but for ObsID:0824030101 of RE J1034+396. \label{fig:psd2}}
\end{figure*}

\begin{deluxetable*}{cccccccc}[h]
	\tablenum{4}
	\tablecaption{The confirmed QPOs of RE~J1034+396 after posterior distribution. \label{tab:result}}
	\tablewidth{0pt}
	\tablehead{ \colhead{} & \multicolumn{3}{c}{ObsID:0506440101} & & \multicolumn{3}{c}{ObsID:0824030101} \\
		\colhead{Time bin} & \colhead{QPO} & & Q & & \colhead{QPO} & & Q
	}
	\startdata
	100 s & $\sim3855^{+367}_{-273}$ & & $\sim33$ & & $\sim3530^{+32}_{-30}$ & & $\sim11073$ \\ \hline
	200 s & $\sim3841^{+194}_{-210}$ & & $\sim29$ & & $\sim3532^{+40}_{-30}$ & & $\sim2773$ \\
	\enddata
	\tablecomments{The upper and lower error bars are 1$\sigma$ standard deviation and the $Q$ factor is the full width at half maxima (FWHM) of the peak.}
\end{deluxetable*}

\section{DISCUSSIONS AND CONCLUSIONS} \label{sec:disandcon}
In this work, six AGNs with long-duration X-ray data observed by \textit{XMM-Newton} were analyzed to search for QPOs using the GP model \textit{celerite}. Of these, only RE~J1034+396 was found to have highly significant QPOs in two observations. These results conflict with previous reports that all six AGNs exhibit QPOs. However, \citet{2012AA...544A..80G} found that, of 104 AGNs with archival \textit{XMM-Newton} observational data, only RE~J1034+396 showed clear evidence of a QPO. Our results ars consistent with their findings.

Many methods have been used to search for QPO signals. The results may have significant differences \citep{2022arXiv221101894P}. Most reports of QPOs are obtained by using LSP and WWZ methods based on Fourier techniques \citep{2015MNRAS.449..467A,2018ApJ...853..193Z,2020MNRAS.495.3538J}. These methods can analyze any form of signal not just pure monochromatic harmonic. The QPO signal appears as a broad peak in the continuous PSD. In recent years, an alternative approach, the GP model is used to successfully characterize the $\gamma$-ray light curves of blazars, suggesting that this model have strong potential for studying AGN variability \citep{2014ApJ...786..143S,2019ApJ...885...12R,2020ApJ...895..122C}. The GP model avoids the problem of red noise leakage and aliasing through directly fitting the light curve \citep{1982ApJ...261..337D,2002MNRAS.332..231U,2003MNRAS.345.1271V}. It allows us to add more physical parameters to the model to make the fit more physically, such as estimating the density of gravitational-wave posteriors \citet{2021MNRAS.508.2090D} and accounting for uncertainty of gravitational-wave models \citet{2016PhRvD..93f4001M}.

Estimating the significance is very important for searching for the QPO behavior in the power spectrum. The methods used to calculate the significance for QPO signals may lead to overestimation of significance \citep{2019MNRAS.482.1270C}. \citet{2021ApJ...907..105Y} studied the gamma-ray light curves of 27 Fermi blazars, which have possible QPO features, by using the GP method. Only 2 of 27 sources have high significant QPO feature. The estimation of the significance of QPO signal by using GP model and other methods should be investigated more deeply with longer time data.

In the PSD of RE~J1034+396, we found two obvious peaks that imply the existence of a QPO signal. These two QPO signals exhibit a standard Gaussian posterior distribution. The much better distribution for ObsID:0824030101 compared to ObsID:0506440101 implies that the ObsID:0824030101 signal is more reliable than that of ObsID:0506440101. We take the higher confidence result as our final period value, which suggests that the RE J1034+396 source has a QPOs of about $3600~s$ (see Table \ref{tab:result}). Our results are very close to those previously reported by \citep{2020MNRAS.495.3538J}.

The interpretation of QPO generation is still debated. \citet{2015ApJ...798L...5Z} applied the orbital resonance model to explain this periodic phenomenon. However, \citet{2021ApJ...906...92S} argued that the orbital resonance model does not accurately describe this phenomenon.
The hot spot model was referenced to explain the physical mechanism of QPO by assuming that there are one or more hot spots at a certain radius of the accretion disk surrounding SMBH \citep{2003APS..APR.P9012S,2004AIPC..714...40S}. The radiation from hot spots closely matches the peak and frequency positions on the PSD. The location of the hot spot is usually identified as near the innermost stable circular orbit (ISCO), which explains the QPO of RE~J1034+396. According to \citet{2009ApJ...690..216G}, the QPO formula of the hot spot model is given by:
\begin{equation}
	M=\frac{3.23\times10^{4}P}{(r^{\frac{3}{2}}+a)(1+z)}
\end{equation}
where $M=M_{BH}/M_{\sun}$, $z$ is the redshift of the object, $r=R/R_{g}$ is the radial position of the hot spot, and $a$ is the spin of the center of a black hole. In the present case, we took $a=0.998$ \citep{2016AA...594A.102C}, $z=0.042$ \citep{1995MNRAS.276...20P} and $P=P_{QPO}$. Assuming the host spot radial position is near the ISCO, $r=R_{ISCO}=1.2$, we estimate the mass of the central black hole to be $\sim4.8\times10^{7}M_{\sun}$, which essentially agrees with the value $\sim1\times10^{7}M_{\sun}$ proposed by \citep{2016AA...594A.102C}. So we might think that this is the physical mechanism that causes the QPO phenomenon. However, that model also has disadvantages: for example, viscous shearing of the disk can decrease survival time of hot spots \citep{2000astro.ph..9169M,2004ApJ...606.1098S}. Therefore, the applicability of this model to this phenomenon needs to be confirmed by further research. We also provide a discussion of other possible physical mechanisms, such as accretion models \citep{1993ApJ...411L..91S,1994ApJ...435L.125M,1996ApJ...468..617Y}. This is a dynamical system of transient chaos that does not require too many assumptions. In this scenario, matter is accreted with a prescribed rate at the edge of the accretion disk, which accumulates at the edge of the disk. When the edge density exceeds some threshold, material begins to fall onto the condensed object at the center. The threshold is generally thought to be due to some unspecified instability. In the simulation, these model can reproduce the QPO phenomenon of X-ray binaries well \citep{1993ApJ...411L..91S,1994ApJ...435L.125M}. For AGNs with the same geometric structure but different mass scales, this model may also be a good internal mechanism of QPO phenomenon.

In summary, we have collected all long-duration X-ray observations of six AGNs reported in the literature to have a QPO. The light curves corresponding to the observations was extracted. The new GP model \textit{celerite} was employed to fit the light curves, and the PSDs of the best-fitting models and the posterior distributions of the periods were calculated. Among all the observations, We found suspected QPOs in only two observations of RE J1034+396 ($\sim3600s$), which approximated the previous study \citep{2020MNRAS.495.3538J}. Our QPO results may support the scenario of periodic signal being generated by a random hot spot in geodesic motion or a phenomenological model called the \textit{A Dripping Handrail}. We then estimated the mass of the central SMBH to be $\sim4.8\times10^{7}M_{\sun}$.

Apart from RE~J1034+396, our work does not corroborate the other five AGN QPO signals found by other methods. Possible reasons are that the significance of these signals is not highly enough to be detected by our model, the reported QPOs are affected by noise, or they may be due to an unknown mechanism that reduces the significance of the signal. Whichever is the case, QPOs in AGNs is a very interesting observational phenomenon needing further study. Finally, we emphasize that reliance on the likelihood function GP model \textit{celerite}, a fast and more statistically powerful fitting model than methods based on traditional Fourier transform approaches, is a reasonable method for searching for QPOs. Our study provides a precedent that the GP model \textit{celerite} could be more widely applied to AGNs or other objects with similar structures.

\begin{acknowledgments}
\centerline{Acknowledgments}
We thank the anonymous referee for constructive comments and suggestions. This work is partly supported by the National Science Foundation of China (12263007 and 12233006), the High-level talent support program of Yunnan Province, and the National Key Research \& Development Program (2018YFA0404204). Haoyang Zhang acknowledge the support of the Postgraduate Research and Innovation Foundation of Yunnan University (2021Y335).
This work is based on observations conducted by \textit{XMM-Newton}, an ESA science mission with instruments and contributions directly funded by ESA Member States and the USA (NASA).
\end{acknowledgments}

\vspace{5mm}
\facilities{\textit{XMM-Newton} (EPIC)}
\software{astropy \citep{2013A&A...558A..33A,2018ApJ....156..123A},
          \textit{celerite} \citep{2017AJ....154..220F},
          emcee \citep{2013PASP..125..306F},
          Numpy, Matplotlib.
          }

\bibliography{QPO}{}
\bibliographystyle{aasjournal}
\appendix
\onecolumngrid
\section{Fitting results and PSD of the non-QPO observations} \label{sec:detailsandpsd}
Except for two observations of ObsID:0506440101 and ObsID:0824030101 of the blazar RE J1034+396, the observations of other five blazars and 4 times in RE J1034+396 did not show QPO. In this appendix, we show all the non-QPO results of GP fitting. For each observation, we show the fitting results, the distributions of the standardized residuals, ACFs of residuals and ACFs of the squared residuals of 100~s and 200~s time bins. The left panel shows the fitting results and the right panel is the PSD distribution obtained by our calculation. The diagram forms of the fitting results and PSD are the same as in Figure \ref{fig:fit1} and Figure \ref{fig:psd1}, respectively.

\begin{figure*}[h]
	\figurenum{A1}
	\gridline{\fig{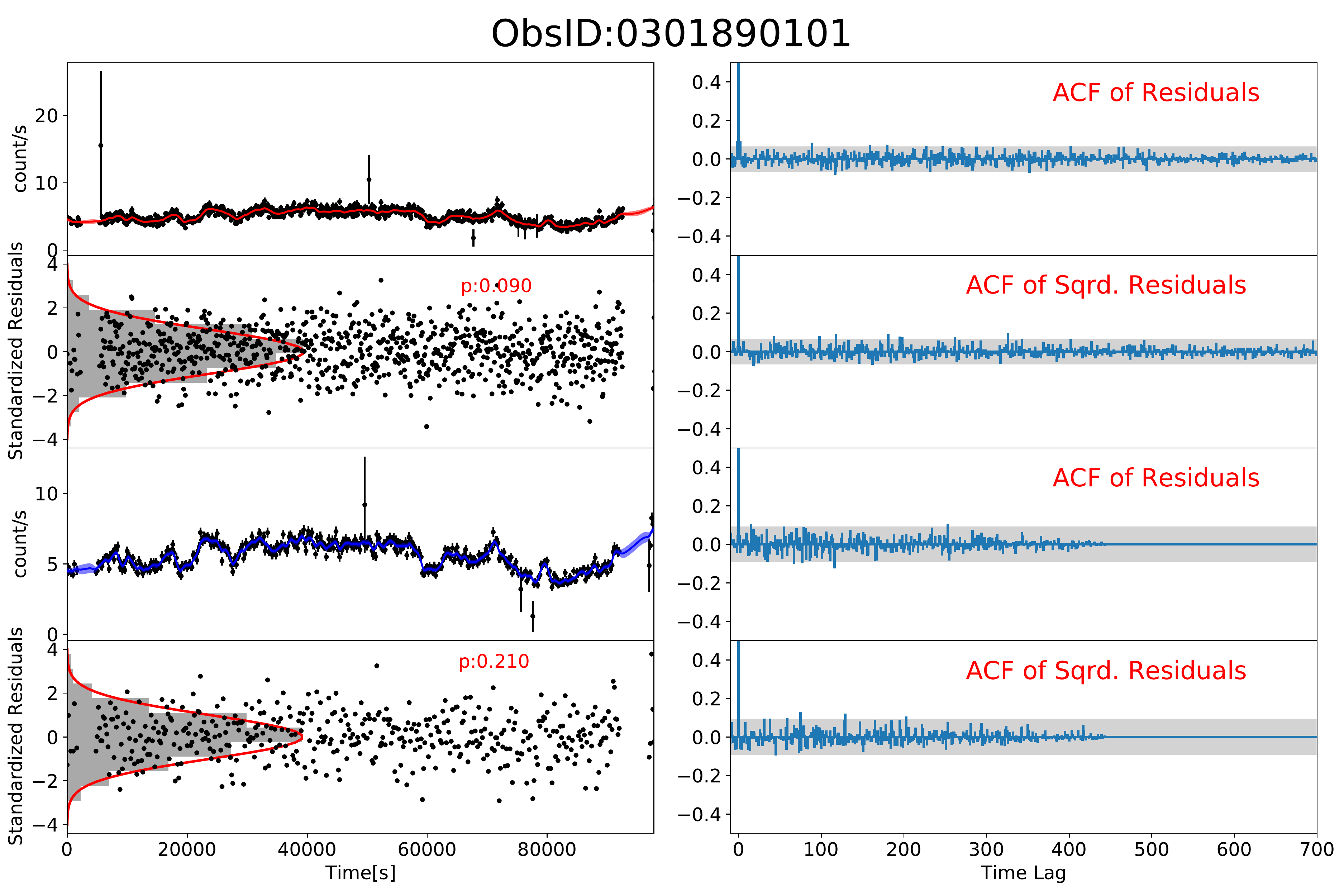}{0.5\textwidth}{}
		\fig{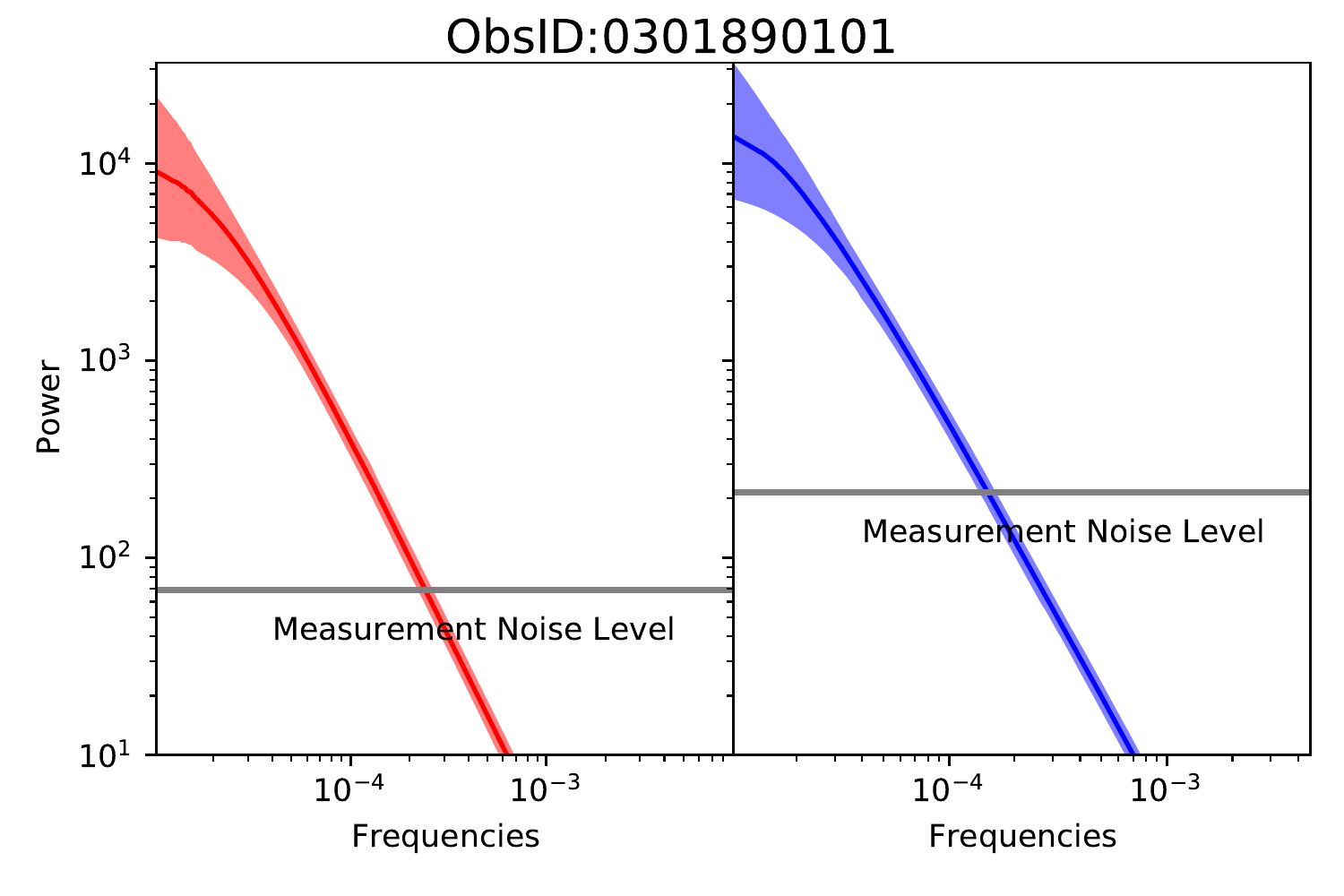}{0.5\textwidth}{}
	}
	\caption{The results of light curve fitting and PSD distribution of ESO 113-G010. The observation number ObsIDs shown in Table \ref{tab:datailsobs} were marked. The left 2 columns show the light curve, the best fitting with 1$\sigma$ deviation, and the distributions of the standardized residuals for 100~s and 200~s time bins respectively. A Gaussian distribution with $\mu=0$ and $\sigma=1$ is shown in first column (red line). The p values are the KS test values. The second column show the ACFs of residuals and the ACFs of the squared residuals for the 100~s and 200~s time bin. The gray area is the 95\% confidence interval of white noise. The PSD distributions with 1$\sigma$ standard deviation were shown in right 2 columns. The red one represents the 100~s time bin, while the blue one is for the 200~s time bin. The gray linea are the measurement noise level.}
\end{figure*}

\begin{figure*}
	\figurenum{A2}
	\gridline{\fig{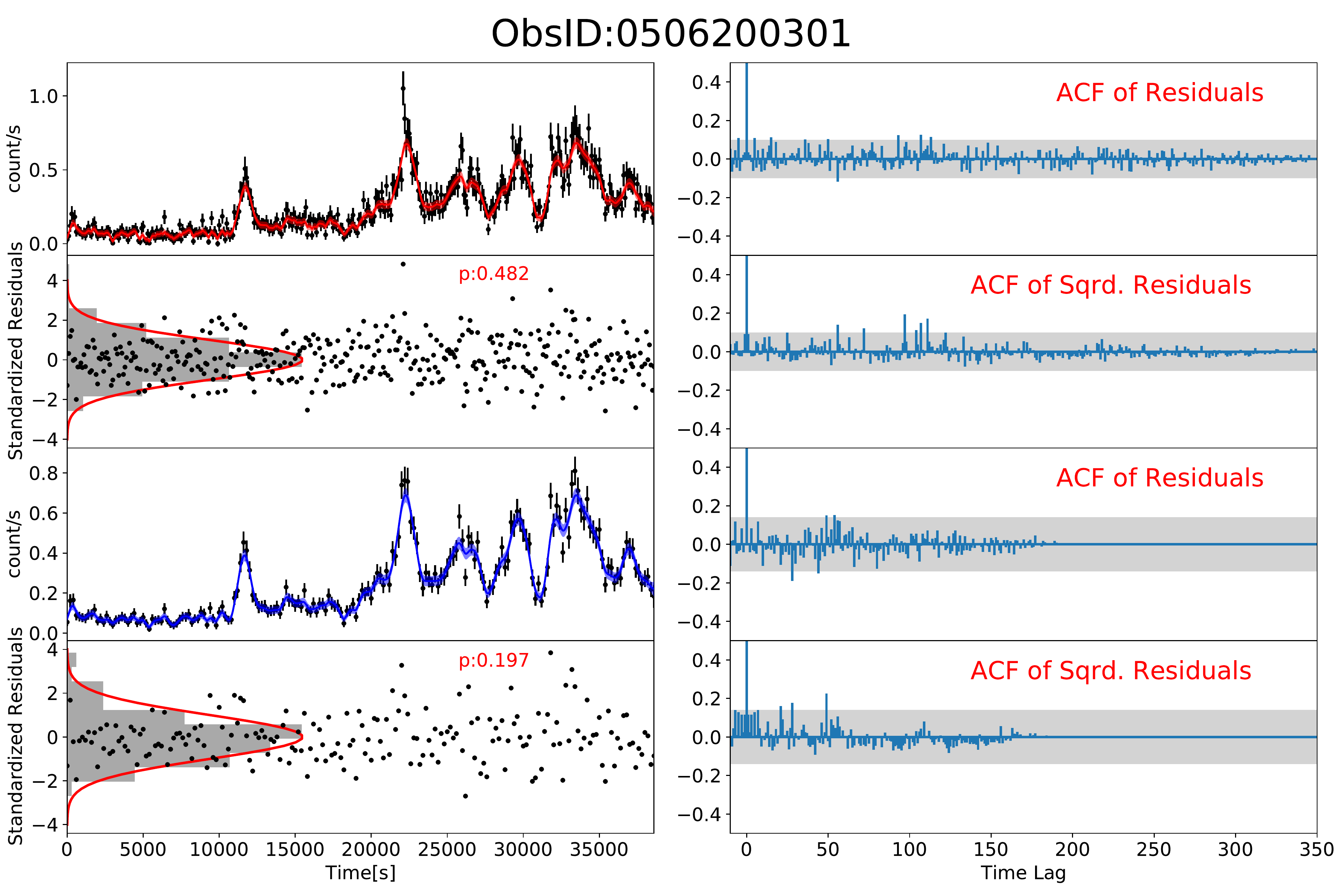}{0.5\textwidth}{}
		\fig{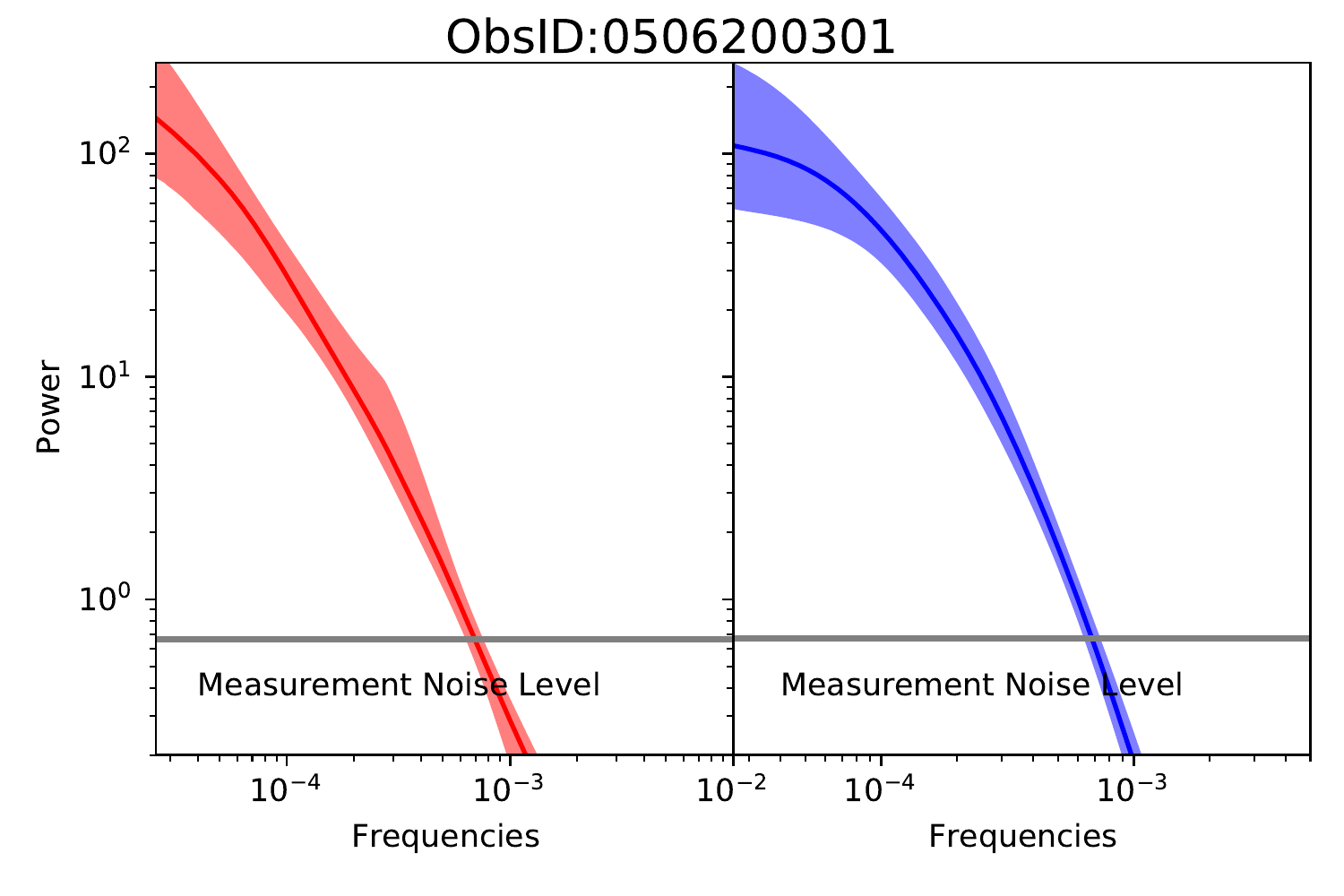}{0.5\textwidth}{}
	}
\end{figure*}

\begin{figure*}
	\figurenum{A2}	
    \gridline{\fig{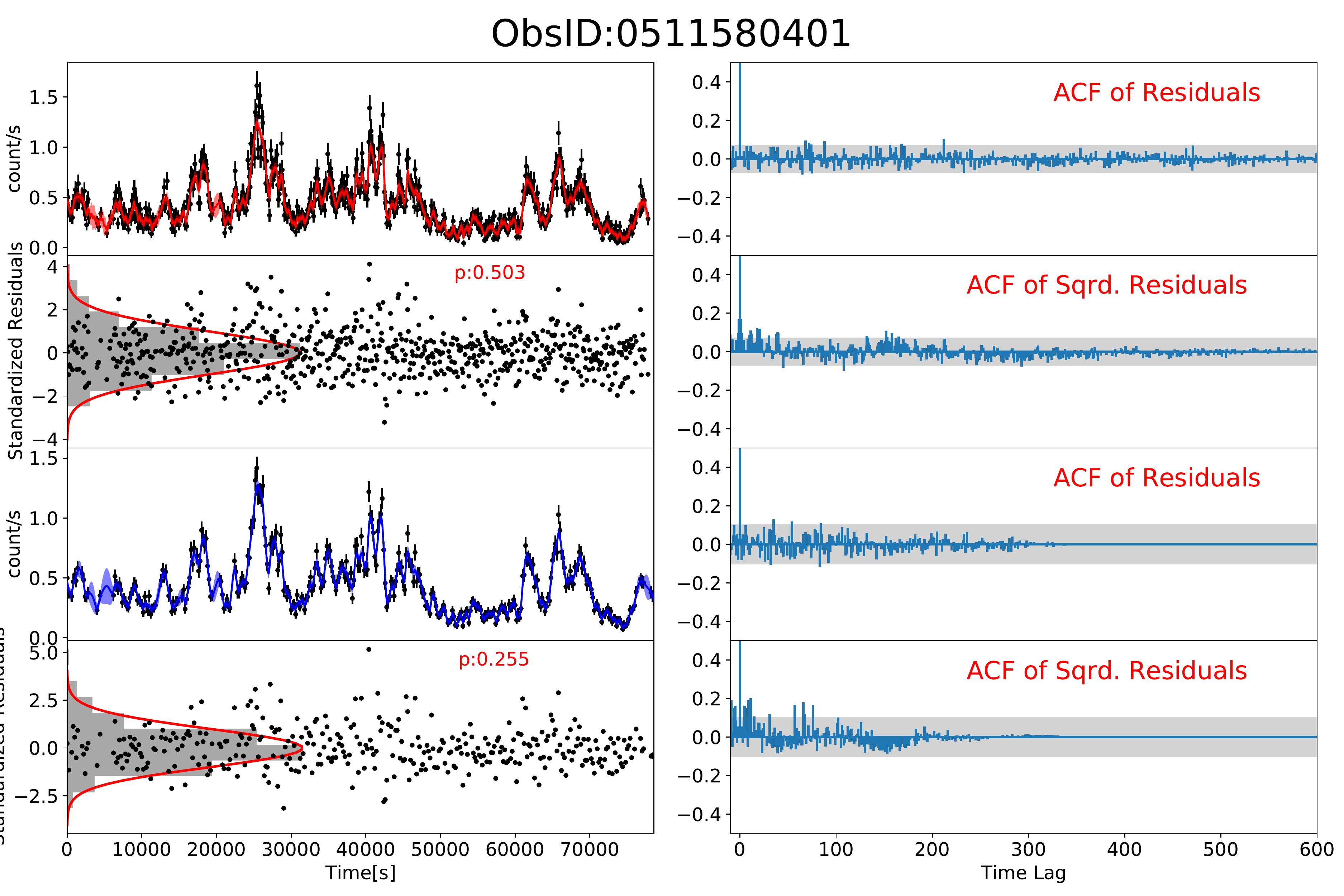}{0.5\textwidth}{}
		\fig{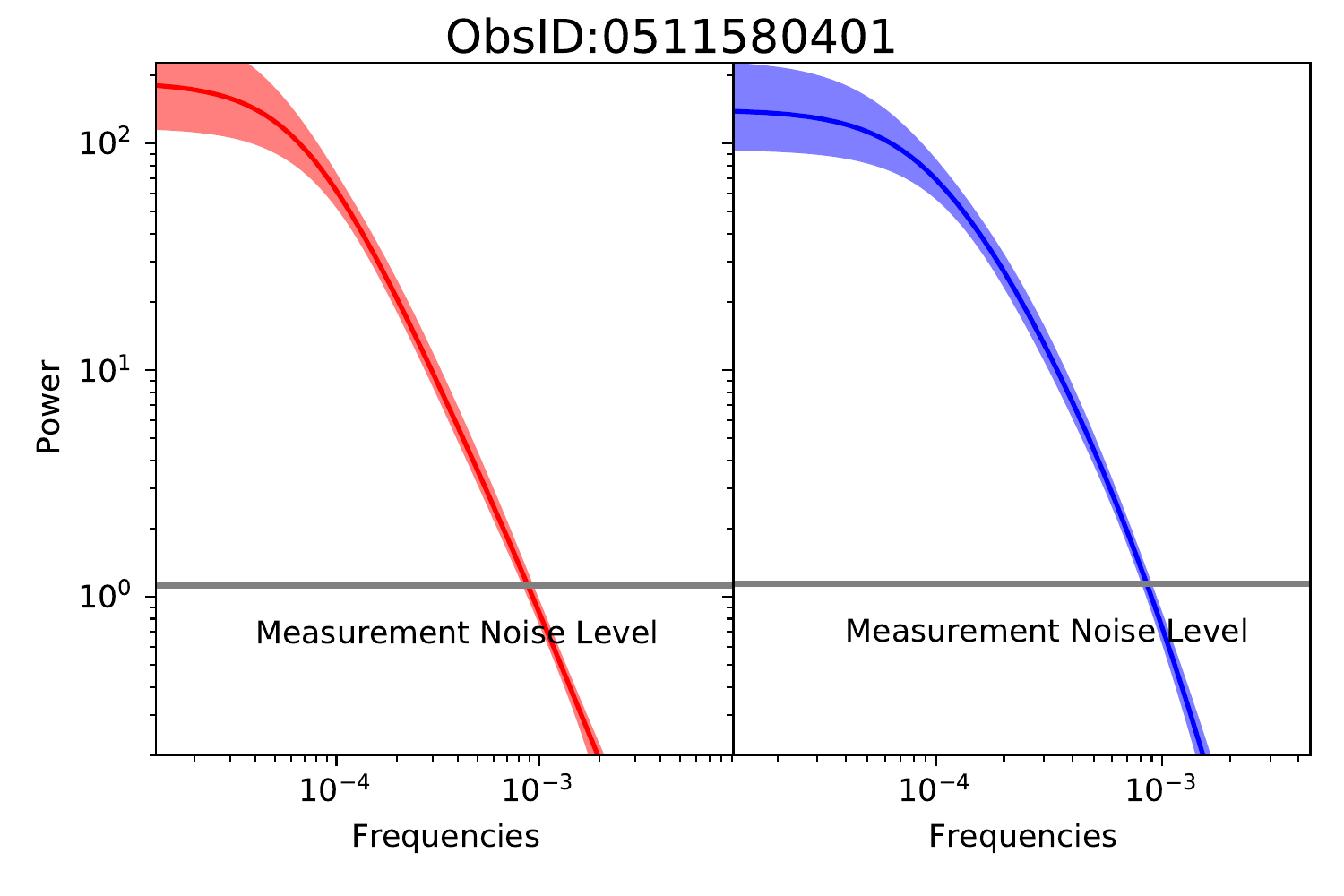}{0.5\textwidth}{}
	}
    \gridline{\fig{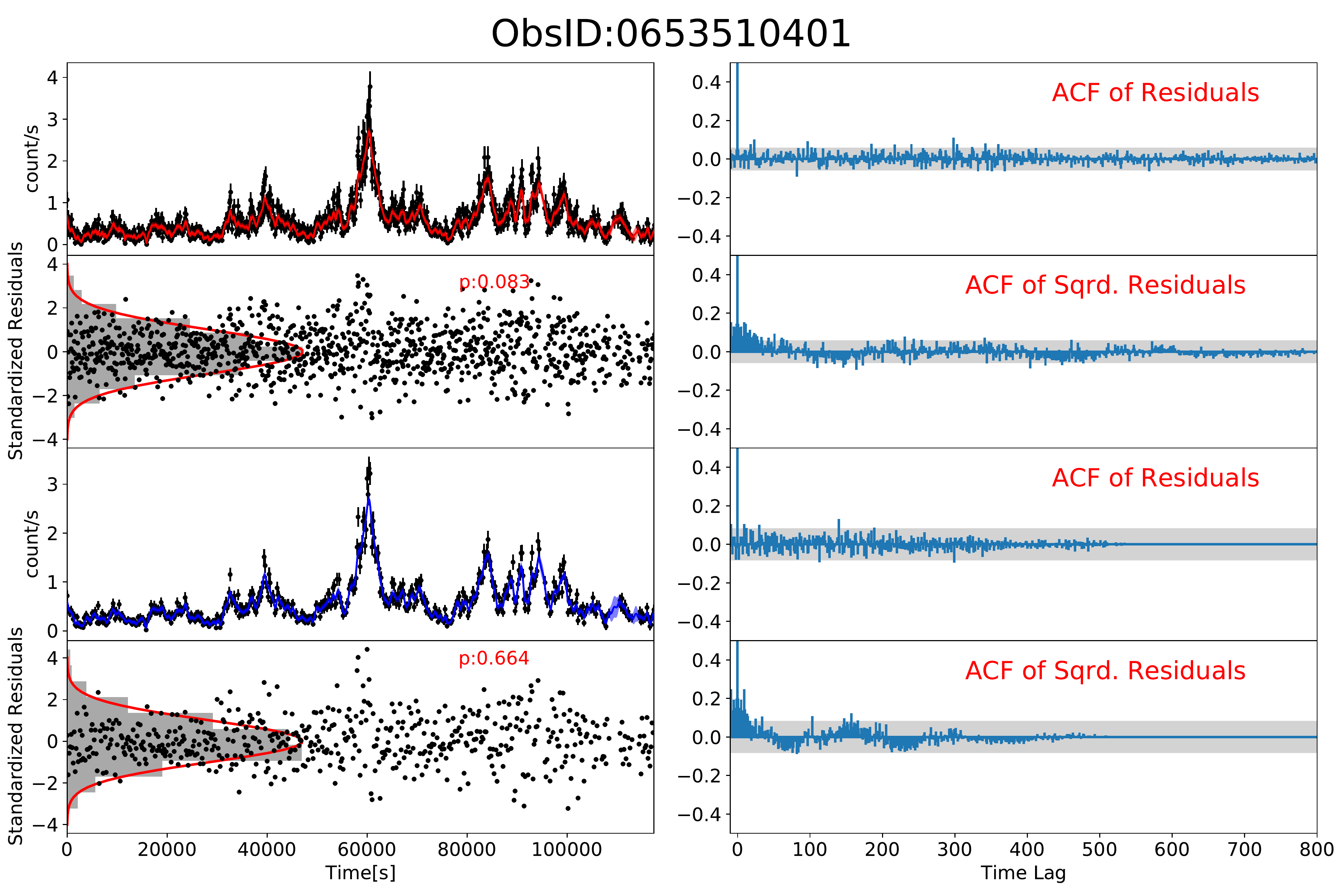}{0.5\textwidth}{}
    	\fig{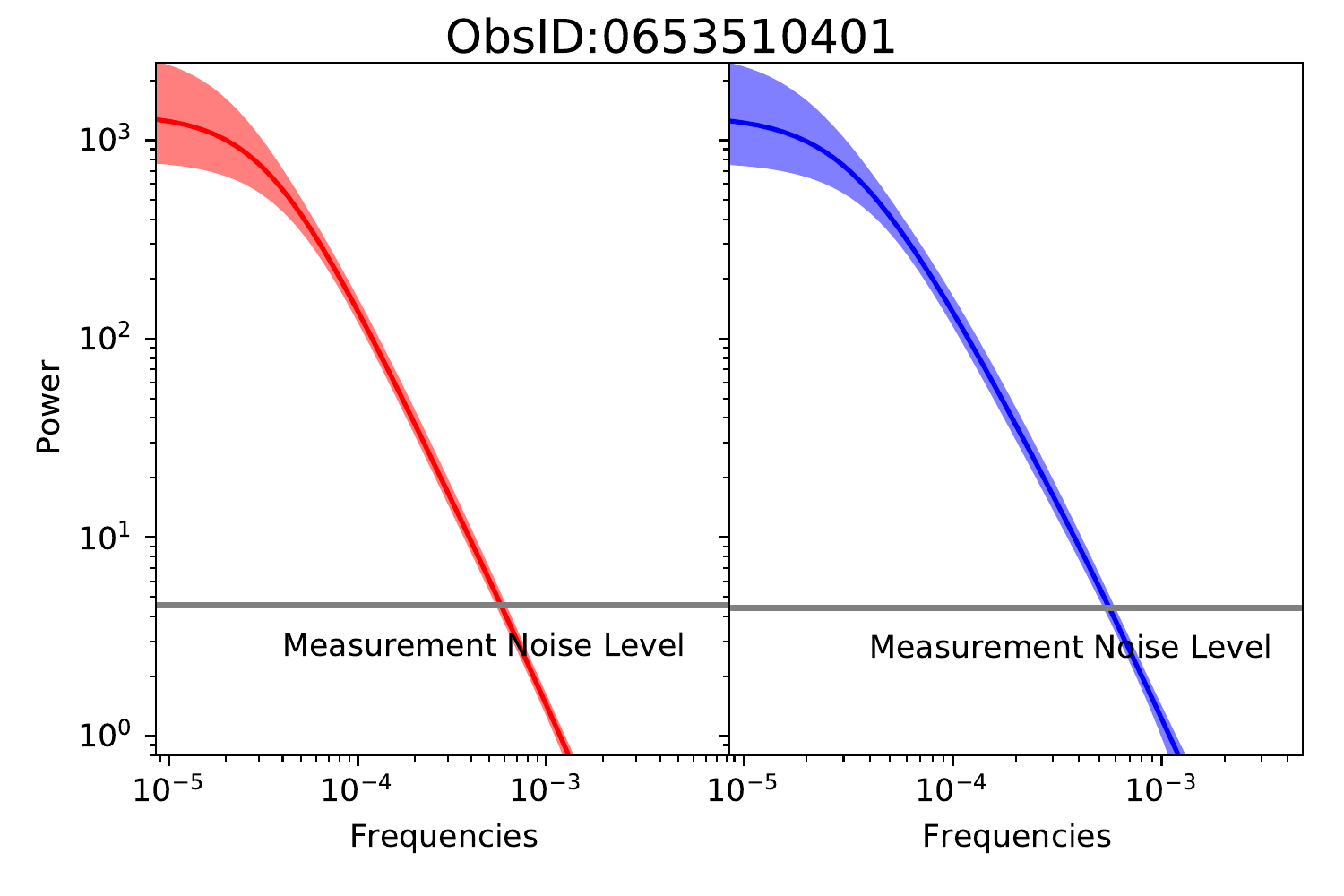}{0.5\textwidth}{}
    }
    \gridline{\fig{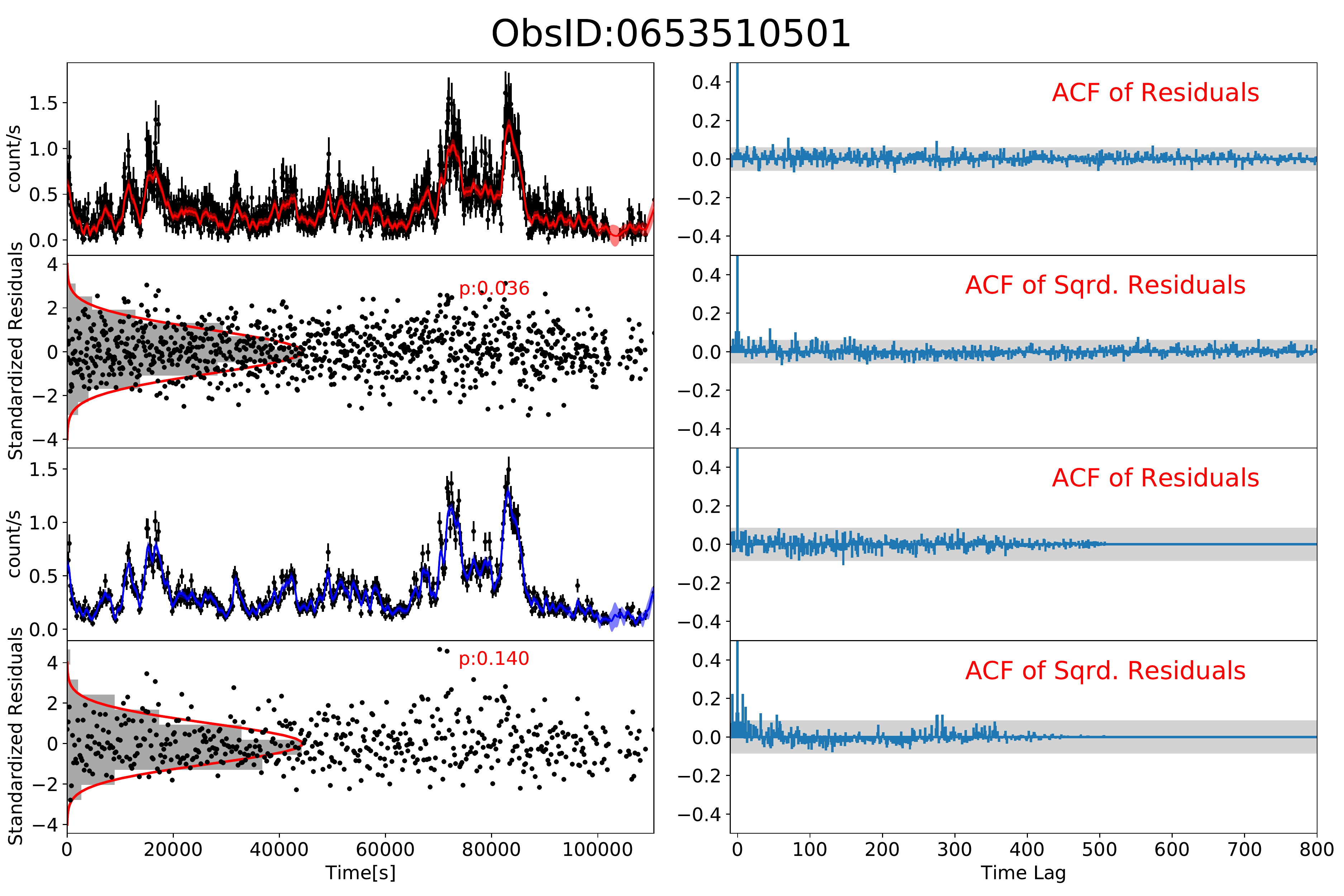}{0.5\textwidth}{}
    	\fig{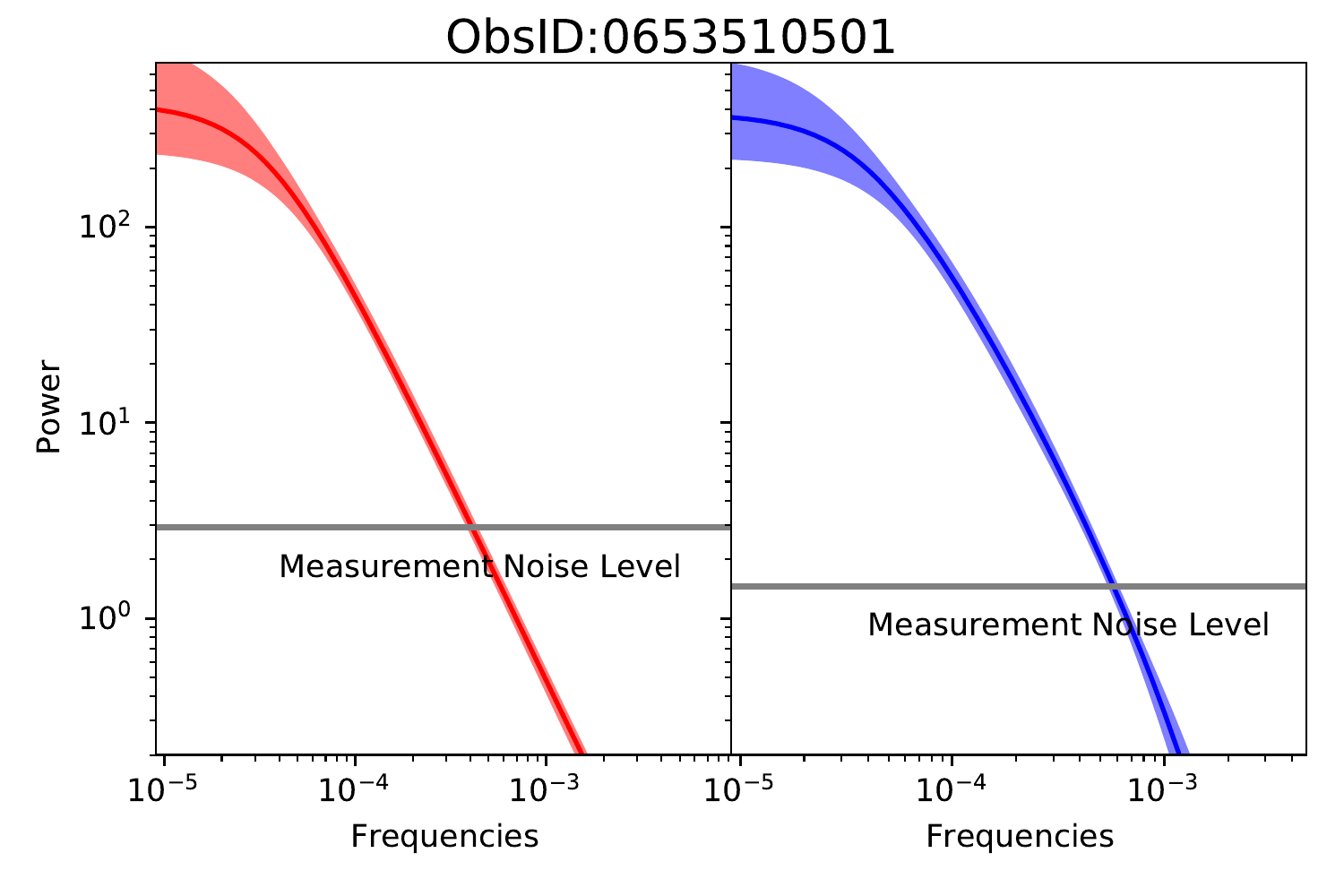}{0.5\textwidth}{}
    }
    \caption{Same as Figure A1, but for 1H0707-495.}
\end{figure*}

\begin{figure*}
	\figurenum{A3}
	\gridline{\fig{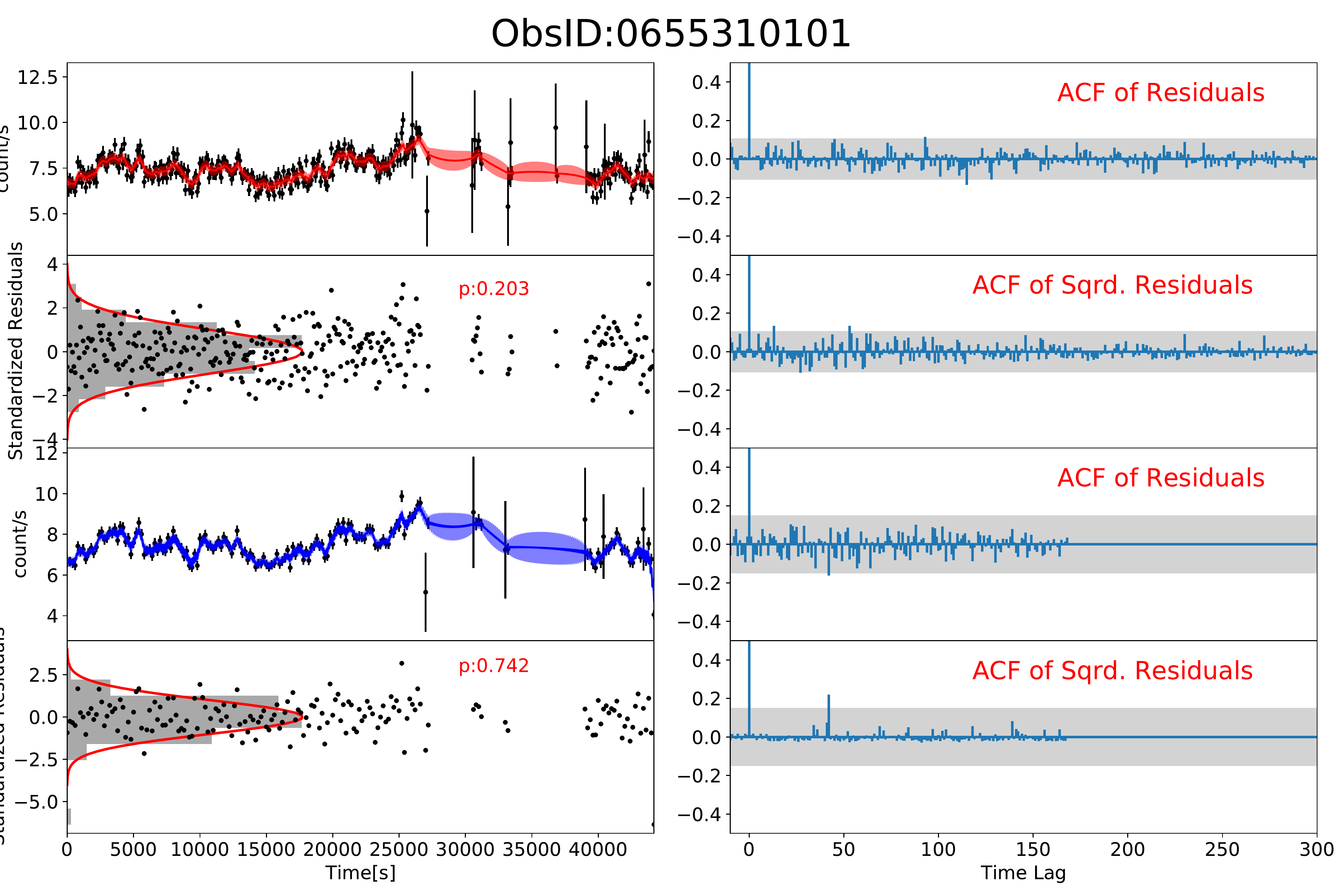}{0.5\textwidth}{}
		\fig{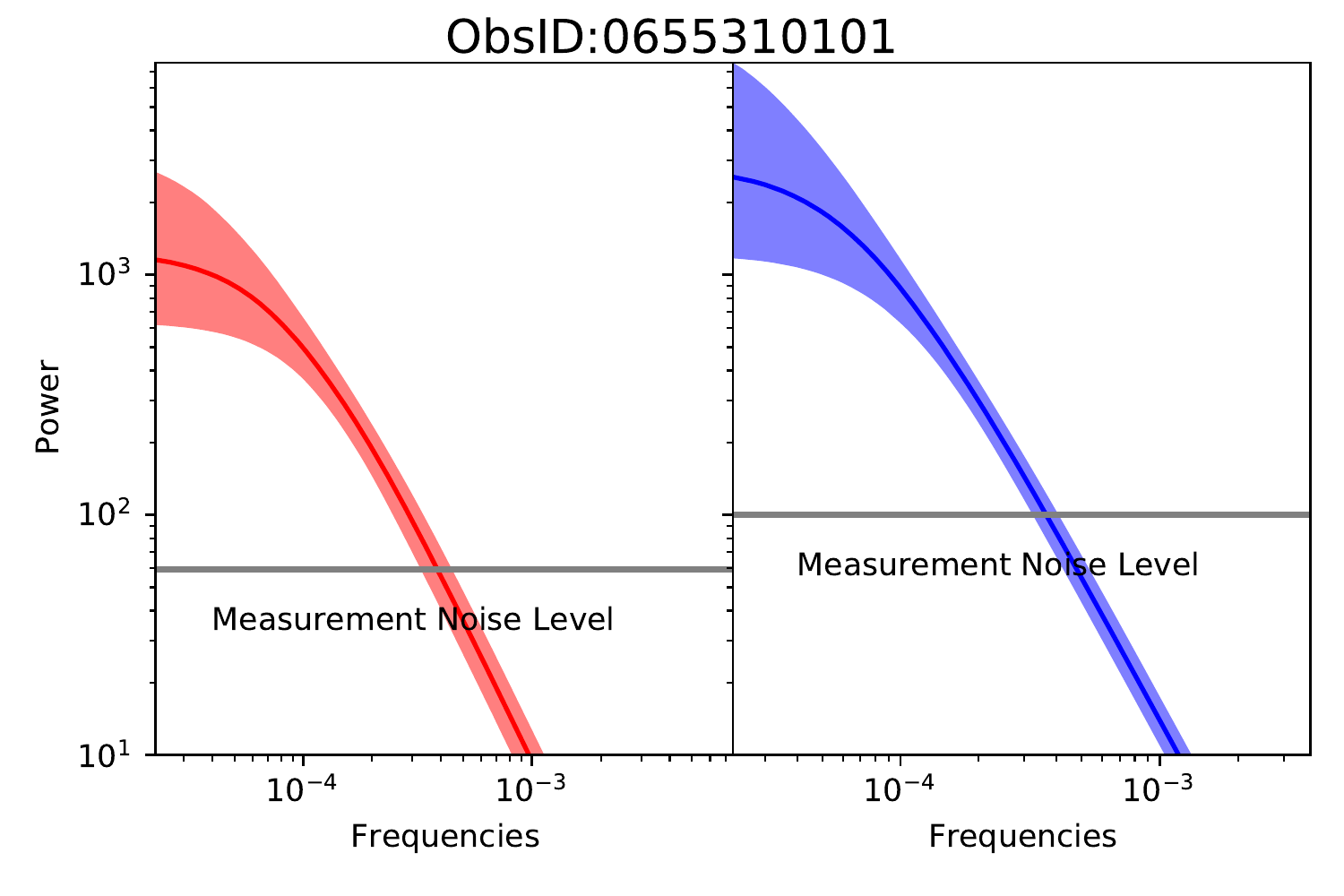}{0.5\textwidth}{}
	}
    \gridline{\fig{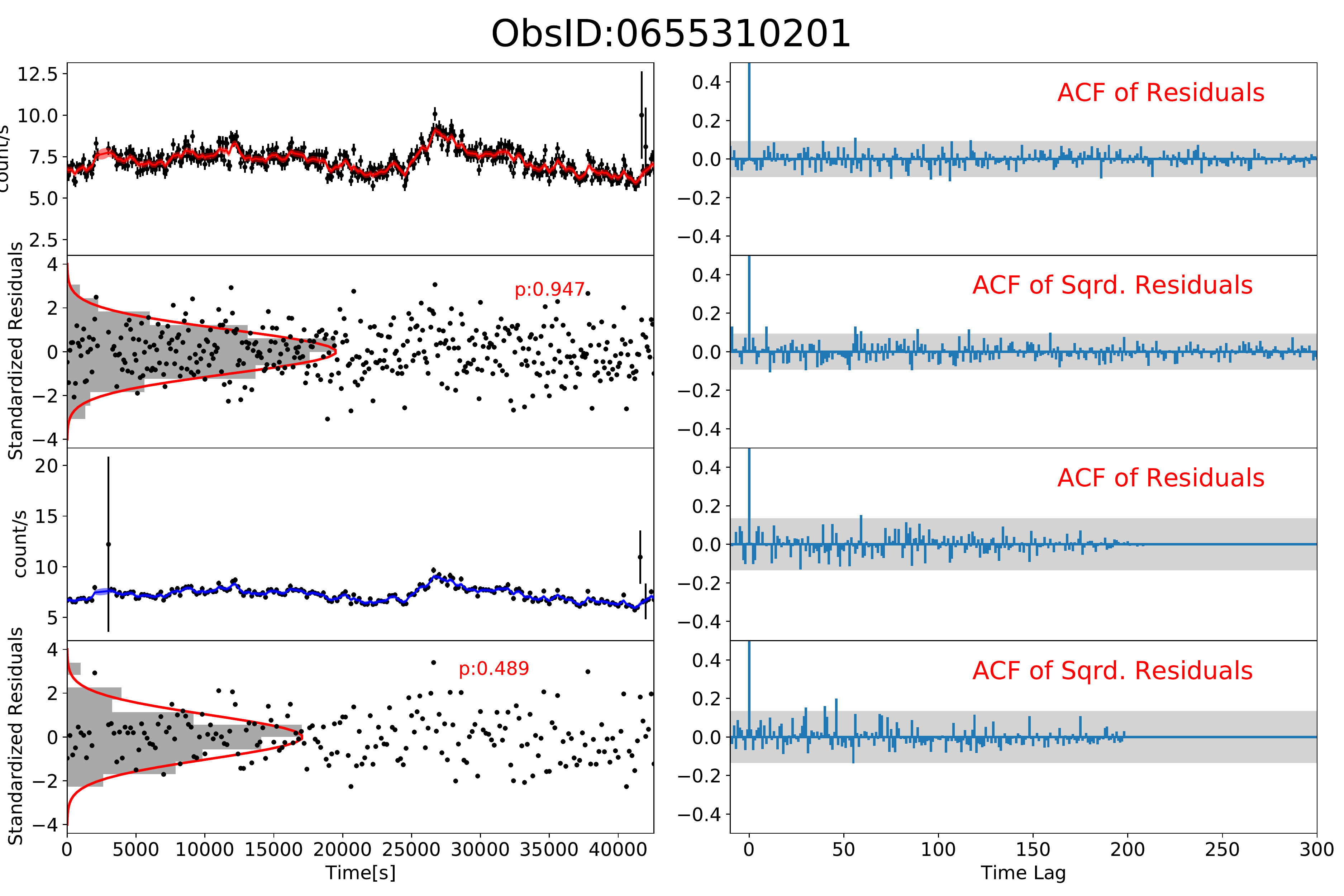}{0.5\textwidth}{}
    	\fig{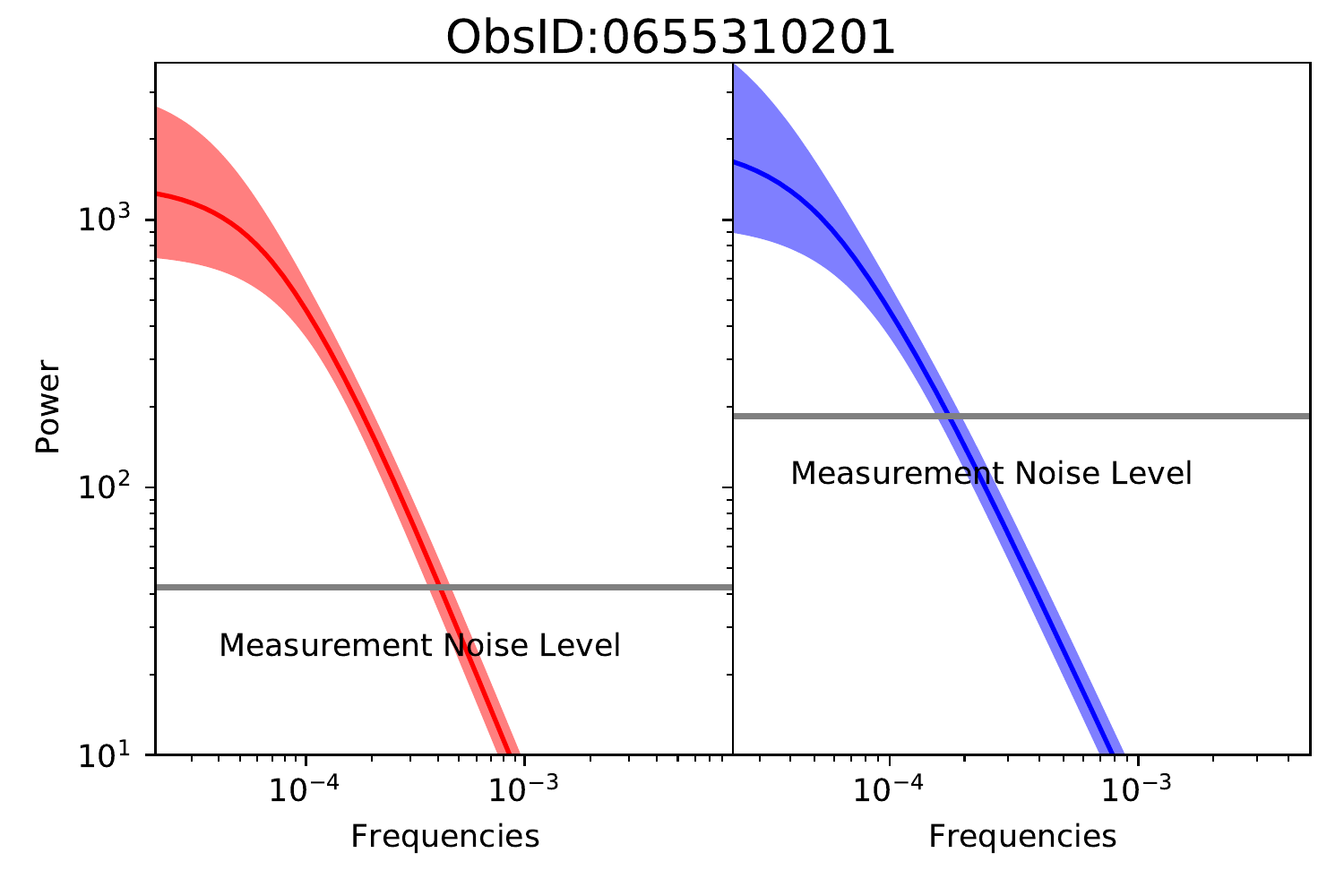}{0.5\textwidth}{}
    }
    \gridline{\fig{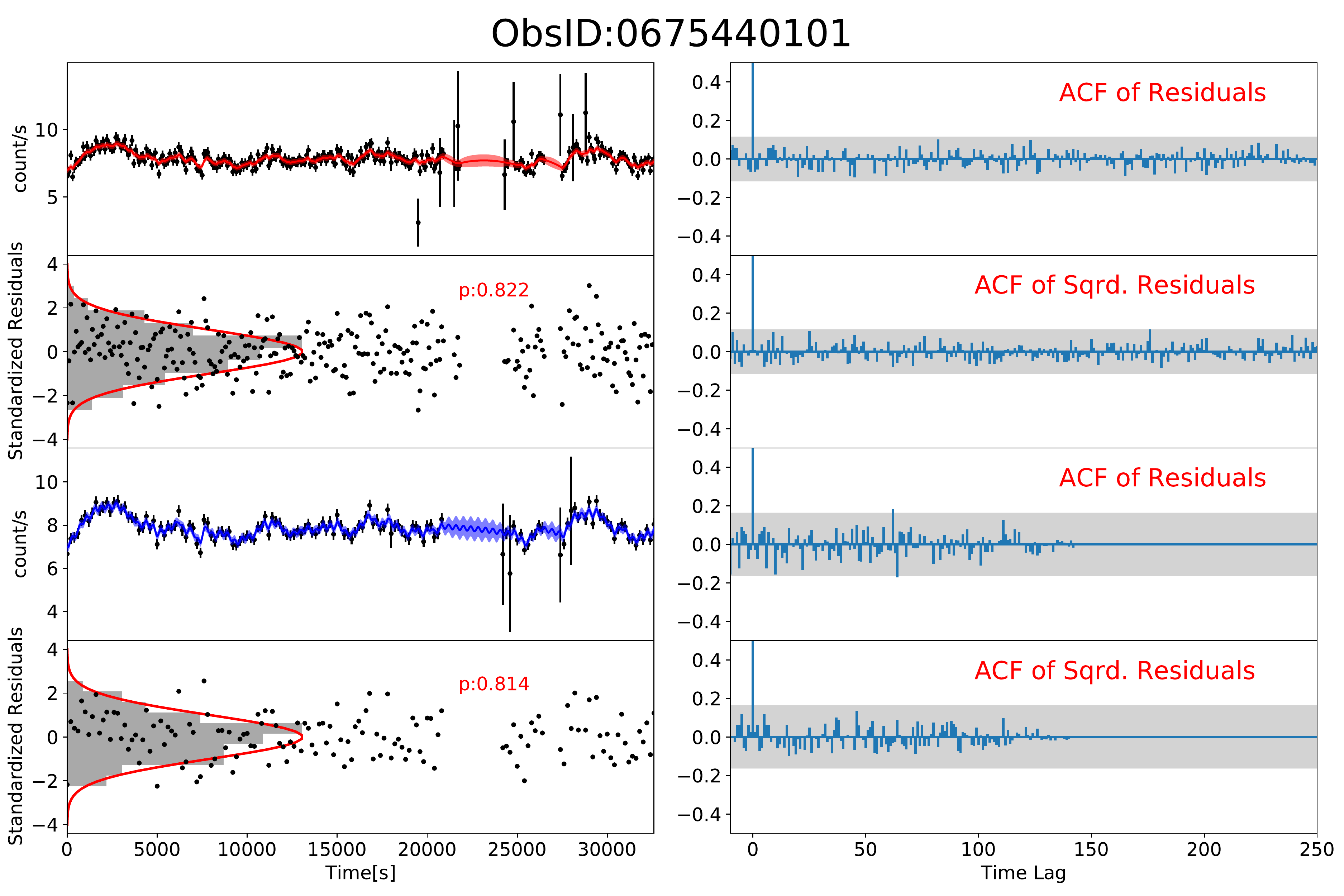}{0.5\textwidth}{}
    	\fig{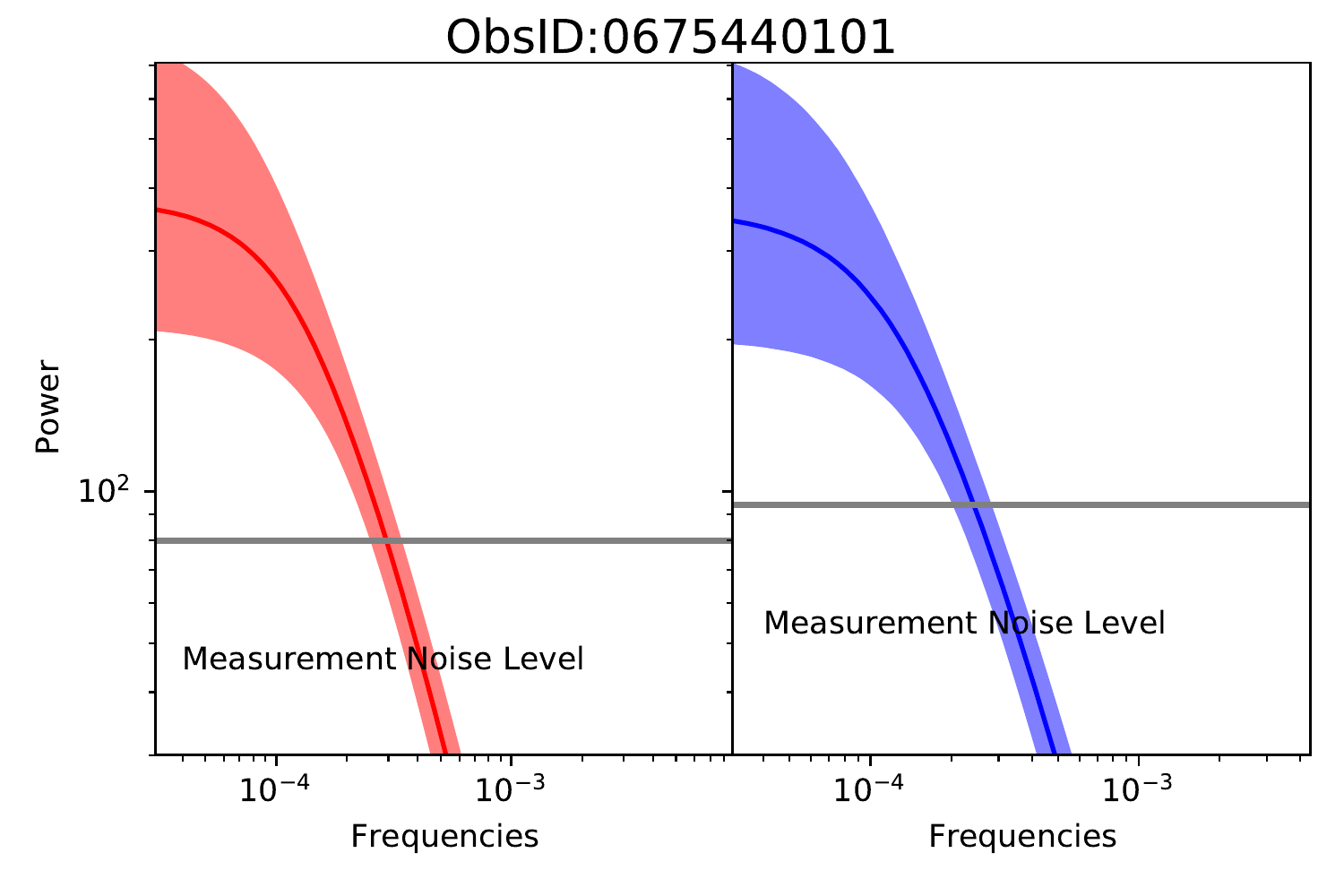}{0.5\textwidth}{}
    }
\end{figure*}

\begin{figure*}
	\figurenum{A3}
    \gridline{\fig{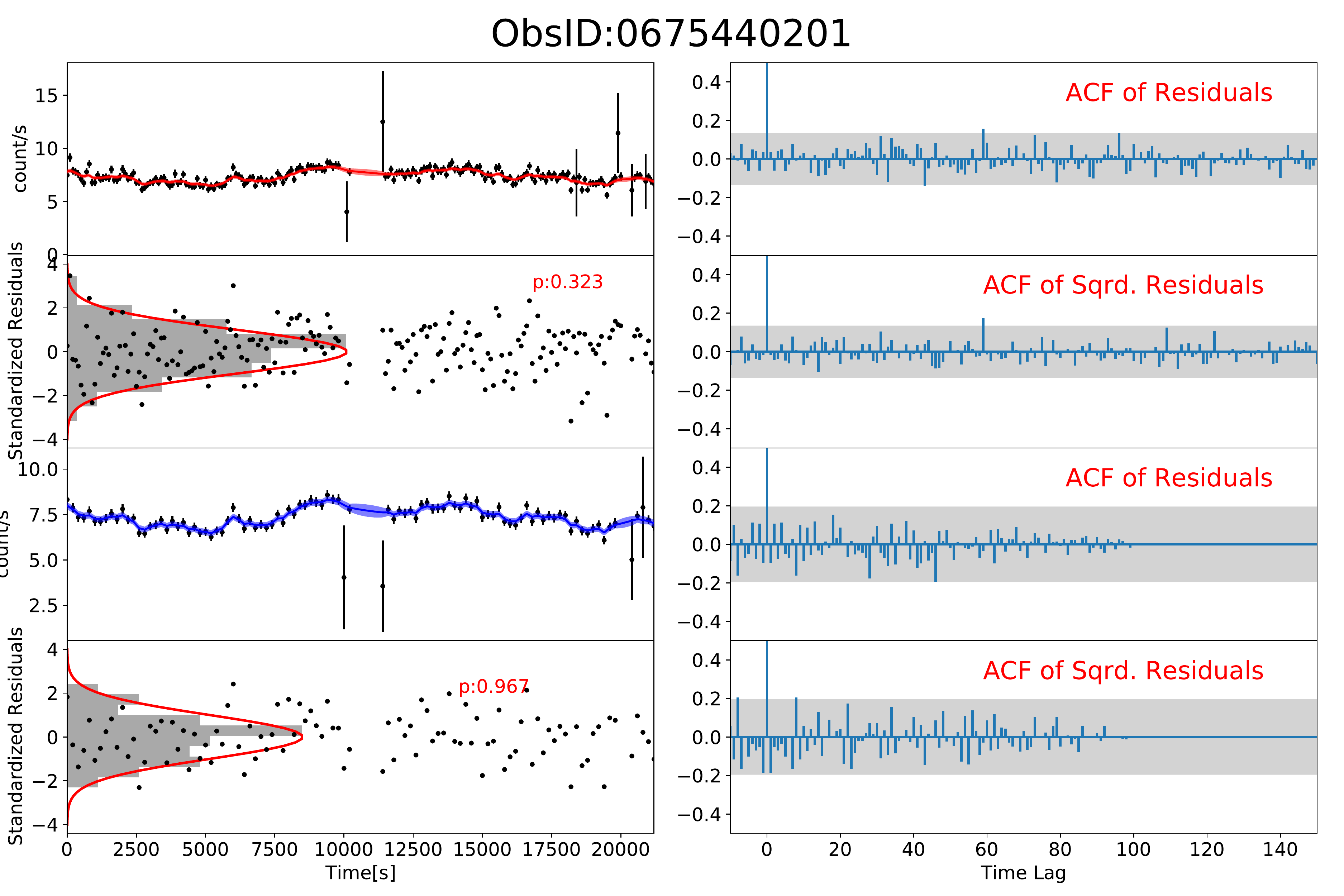}{0.5\textwidth}{}
    	\fig{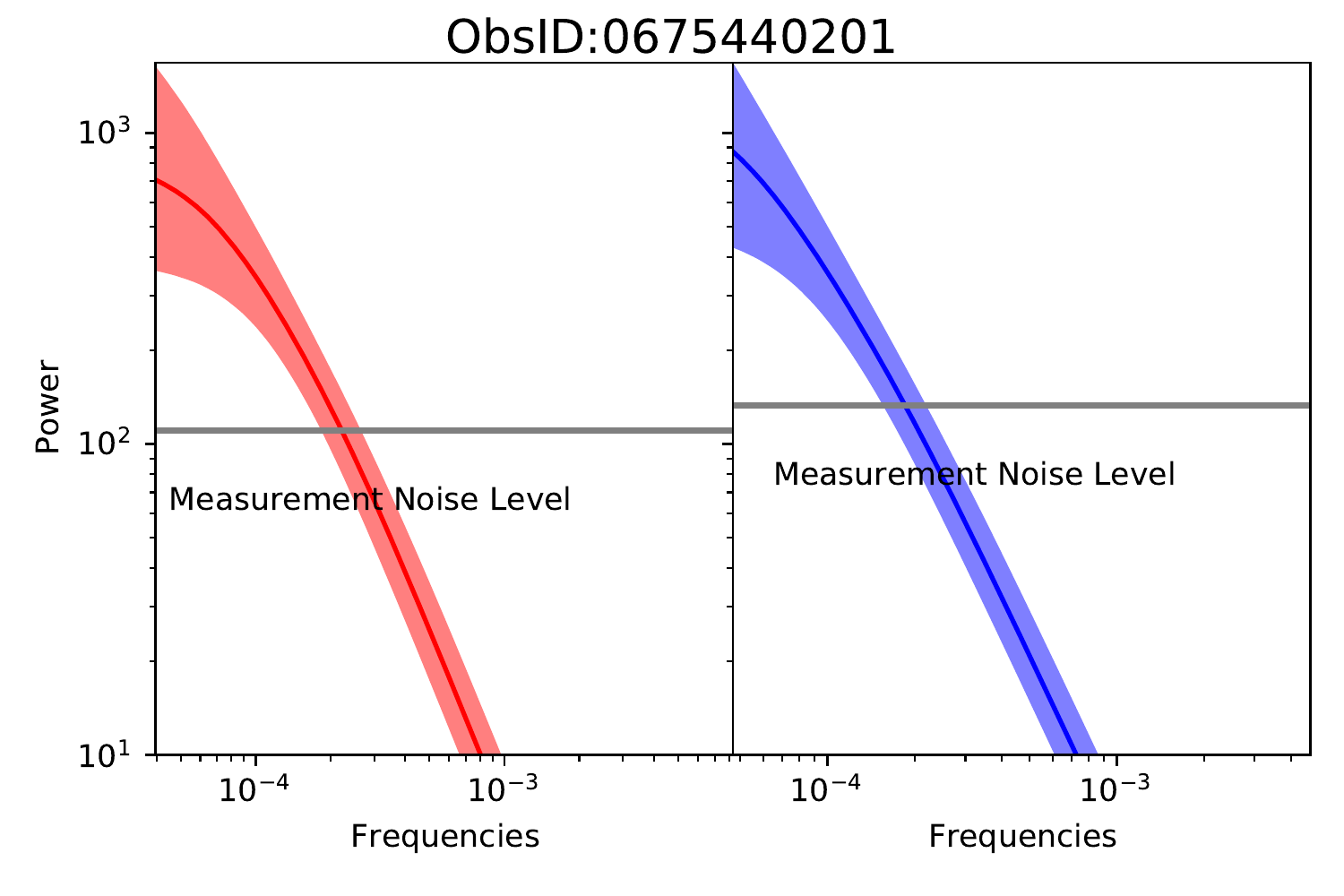}{0.5\textwidth}{}
    }
	\caption{Same as Figure A1, but for RE J1034+396.}
\end{figure*}

\begin{figure*}
	\figurenum{A4}
	\gridline{\fig{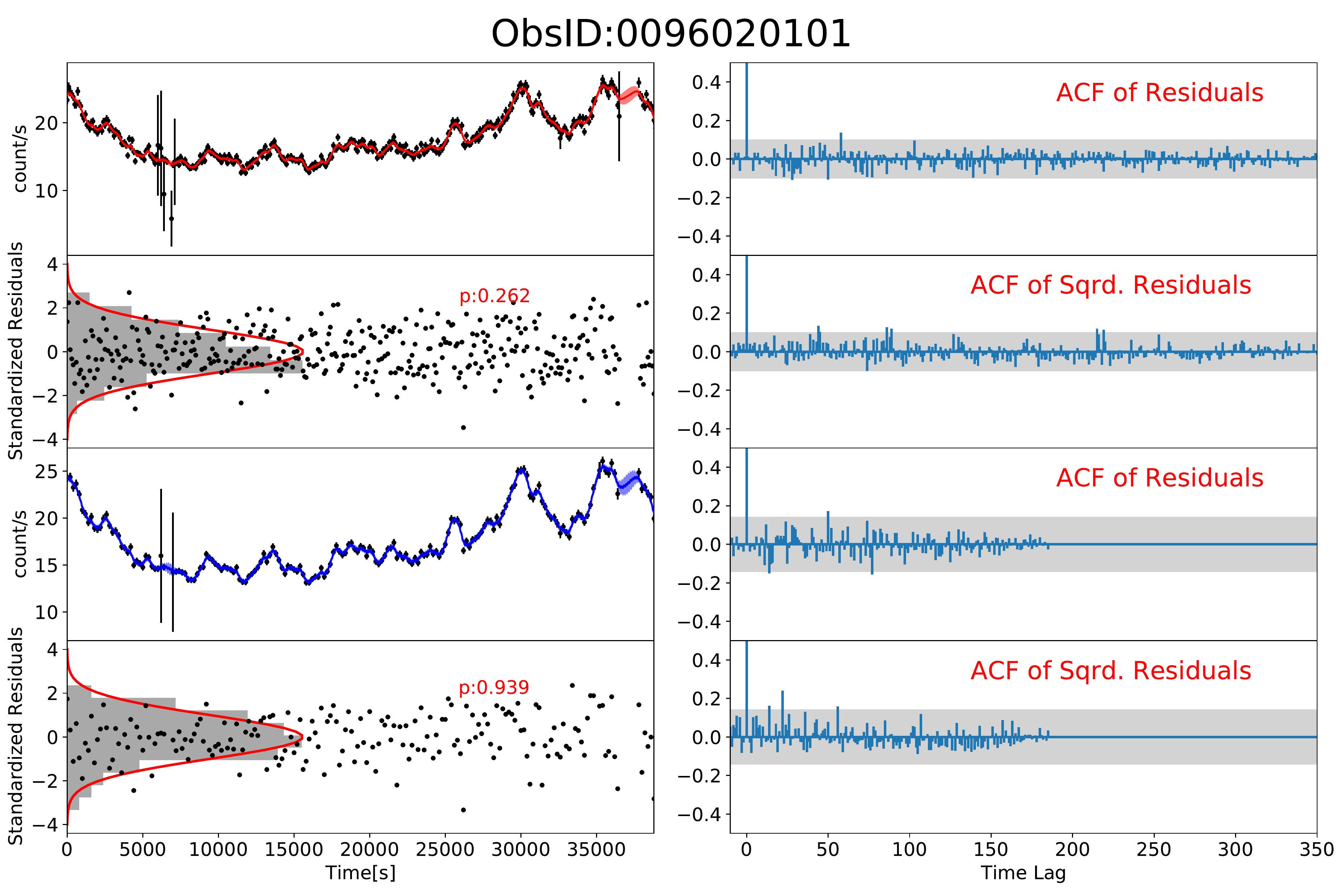}{0.5\textwidth}{}
		\fig{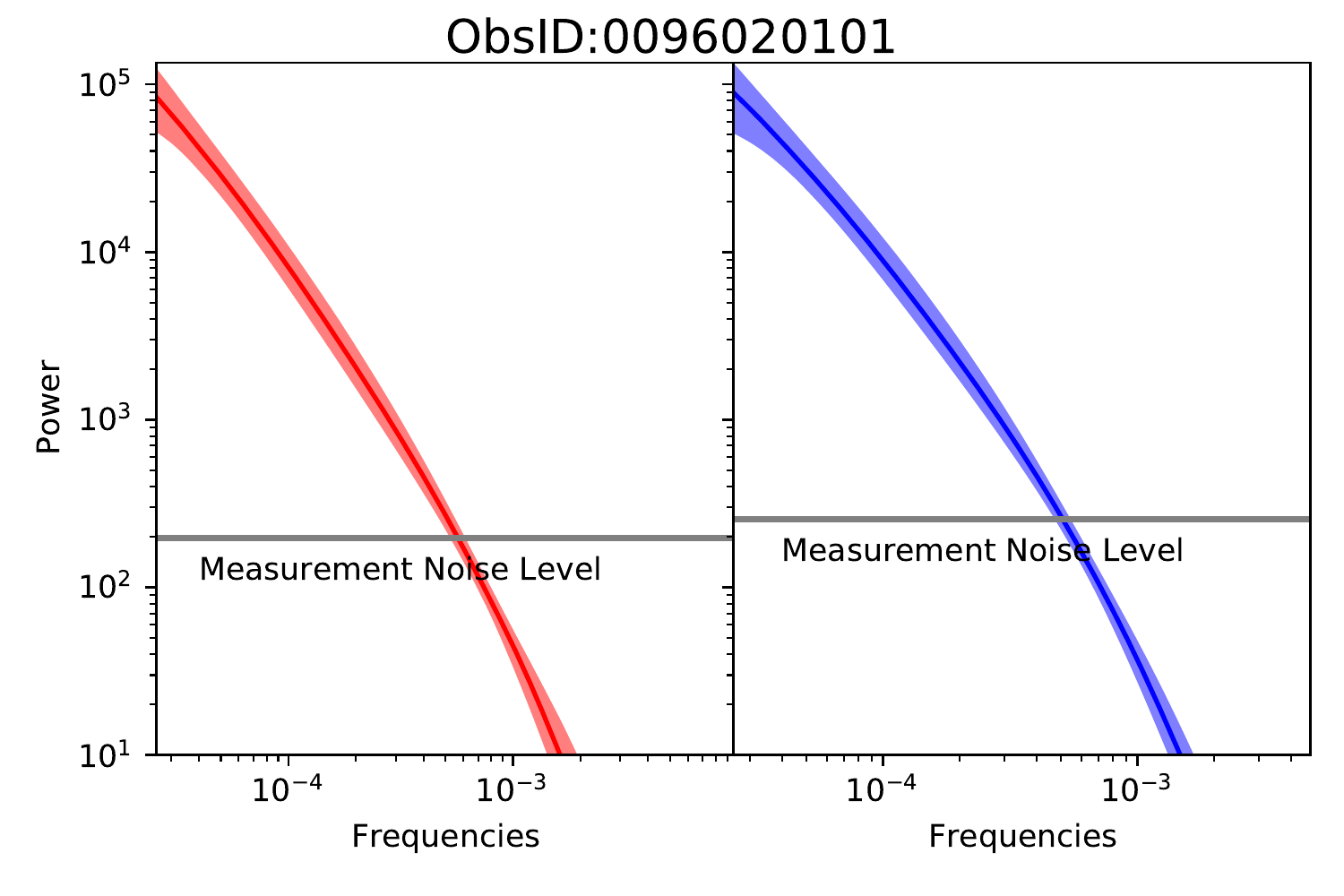}{0.5\textwidth}{}
	}
	\gridline{\fig{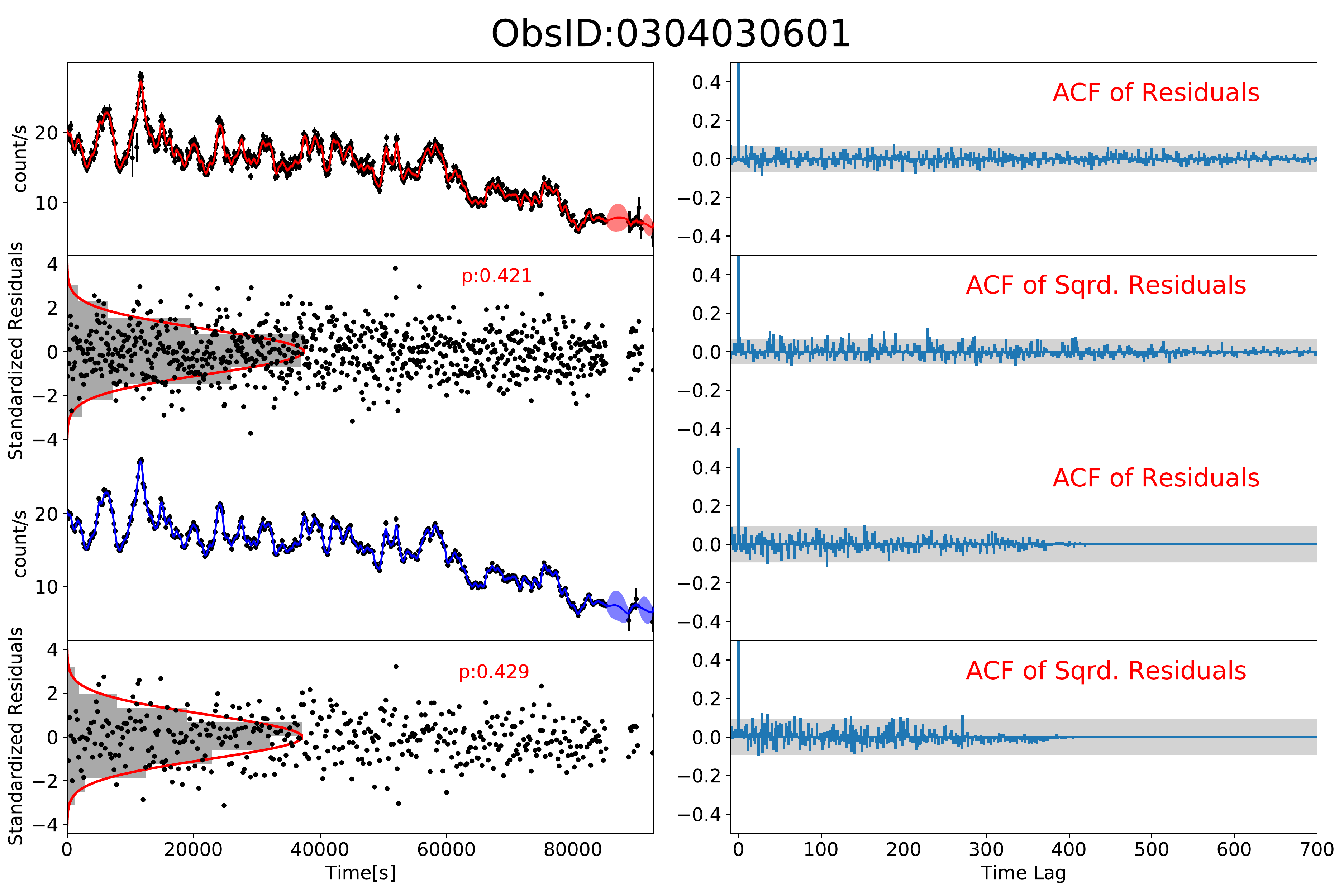}{0.5\textwidth}{}
		\fig{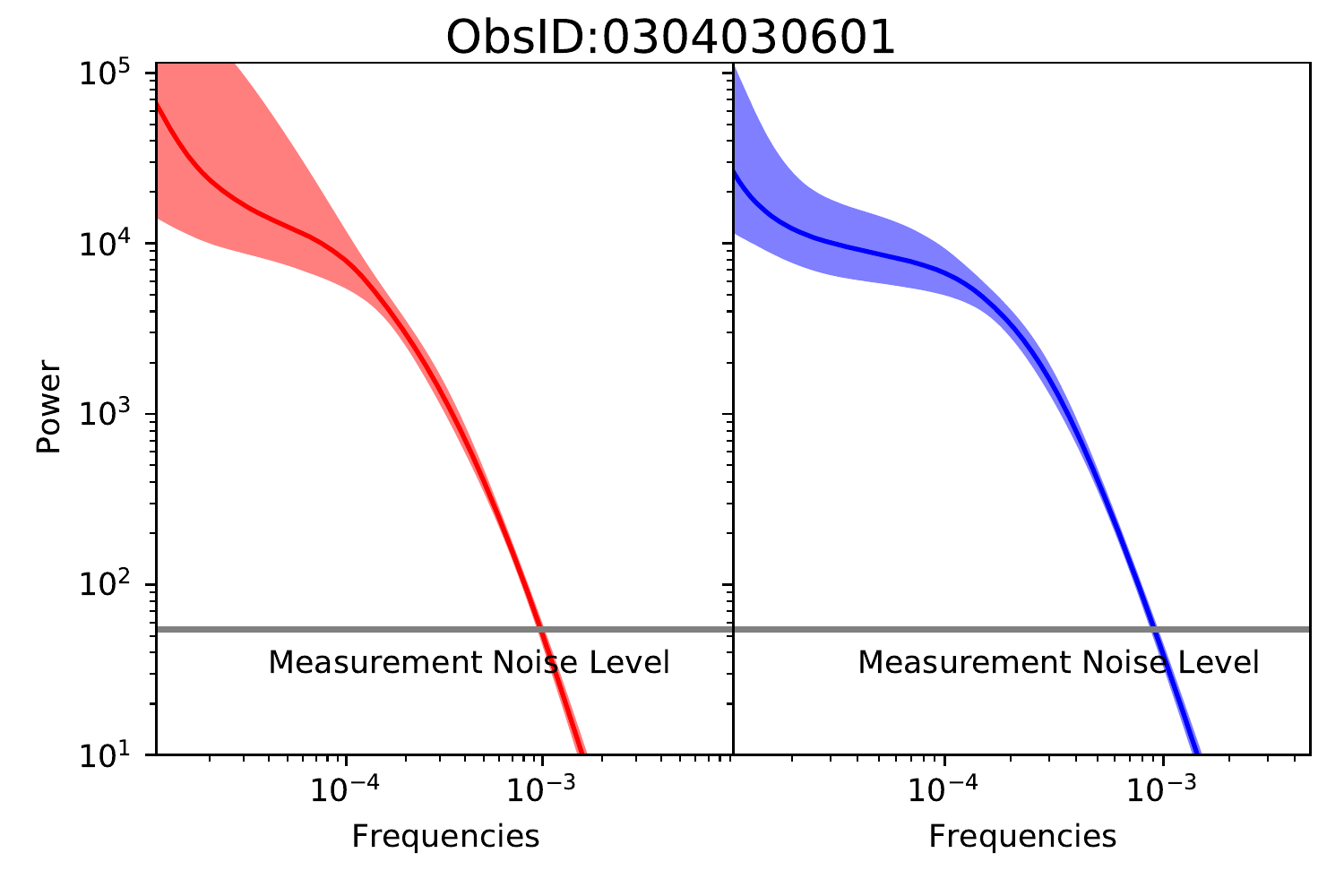}{0.5\textwidth}{}
	}
    \caption{Same as Figure A1, but for Mrk 766.}
\end{figure*}

\begin{figure*}
	\figurenum{A5}
	\gridline{\fig{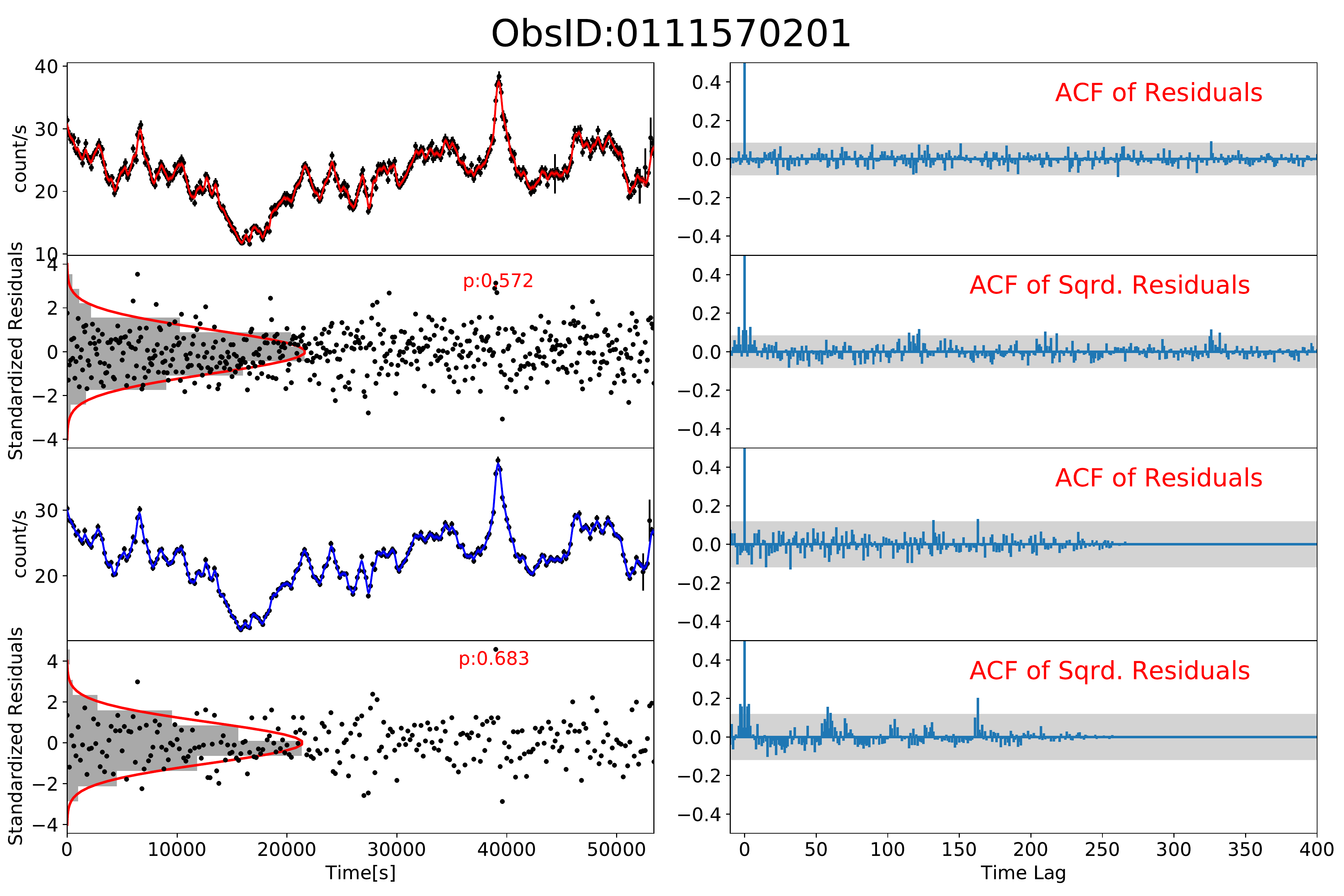}{0.5\textwidth}{}
	\fig{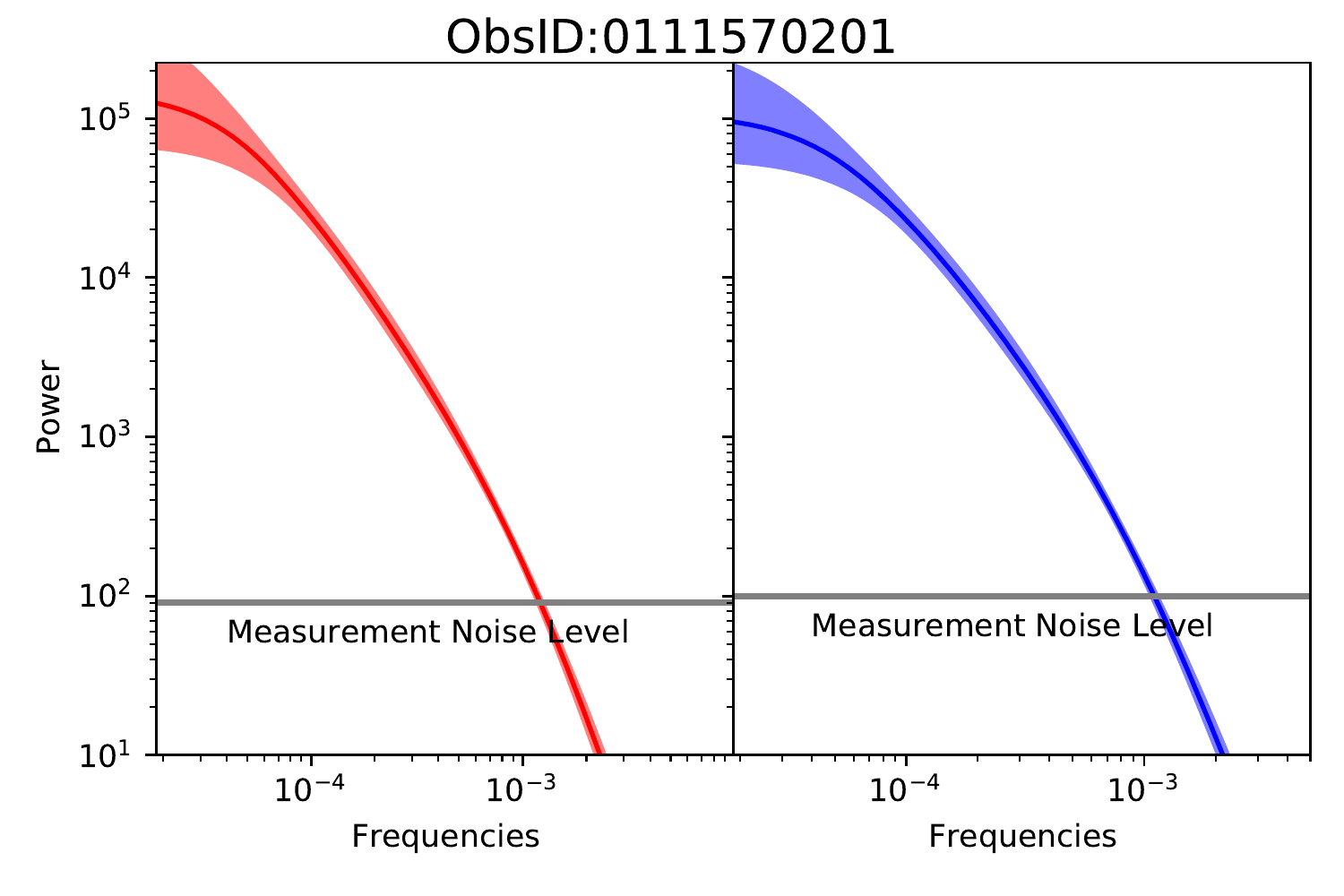}{0.5\textwidth}{}
    }
    \caption{Same as Figure A1, but for MCG-06-30-15.}
\end{figure*}

\begin{figure*}[h]
	\figurenum{A6}
	\gridline{\fig{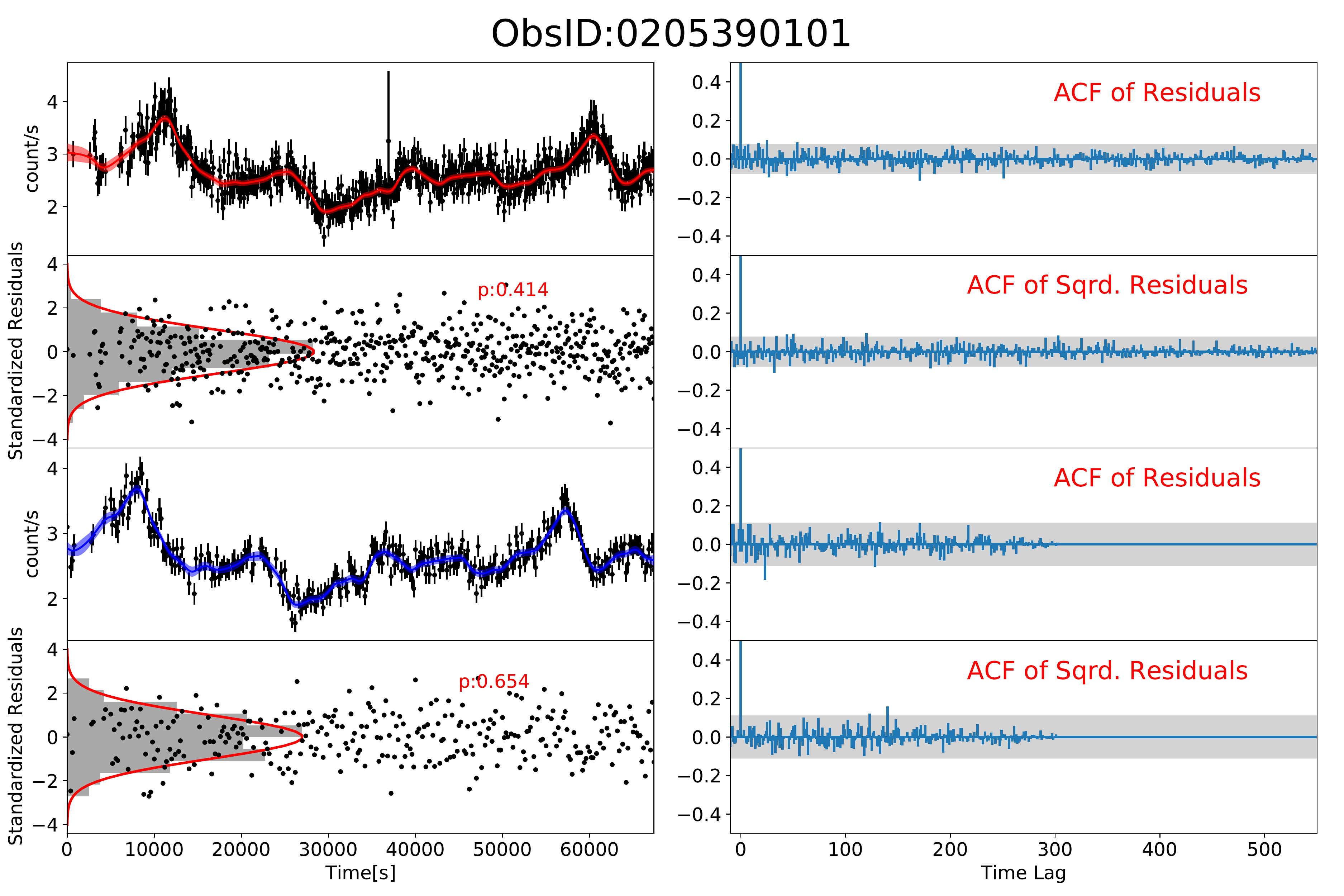}{0.5\textwidth}{}
		\fig{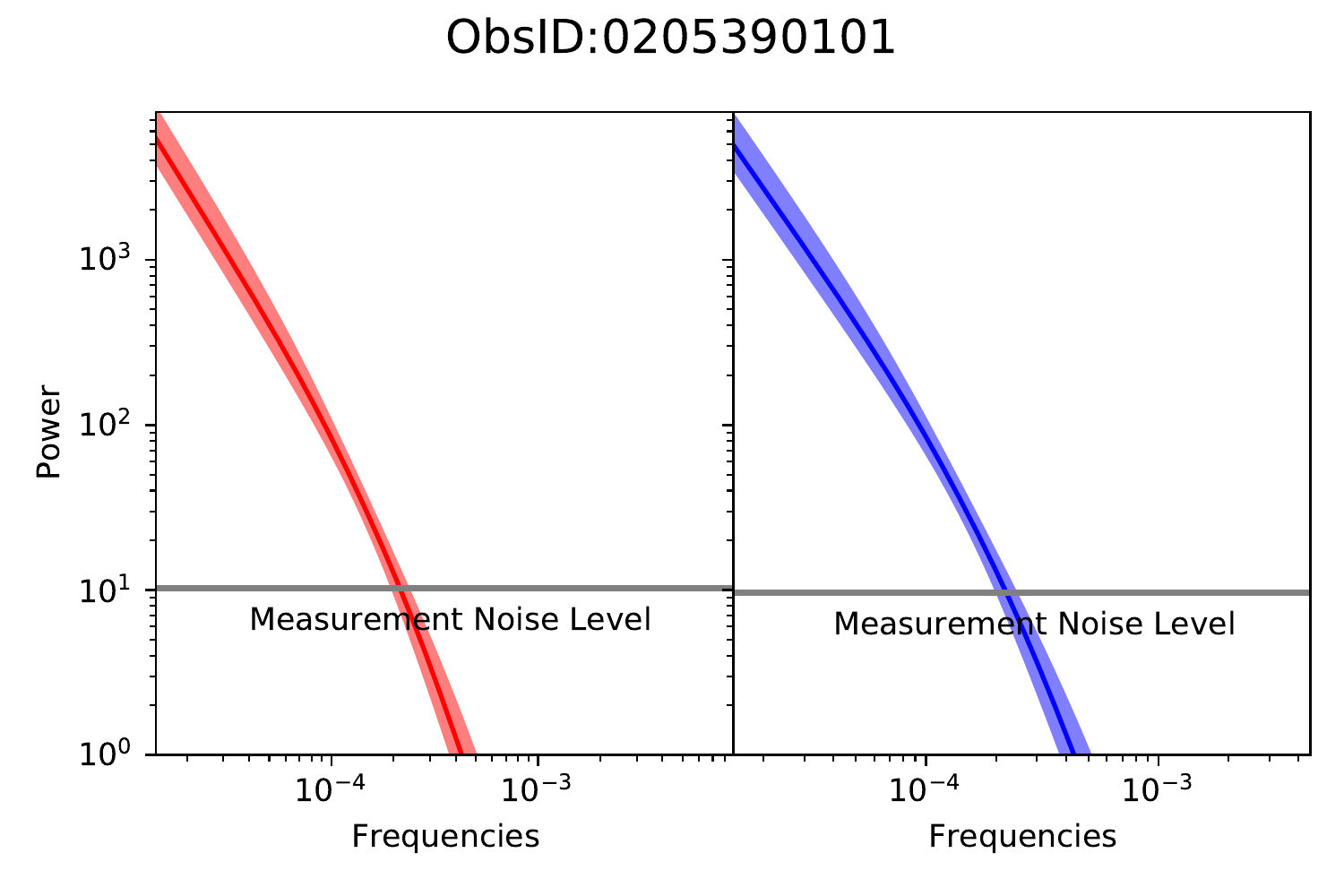}{0.5\textwidth}{}
	}
	\caption{Same as Figure A1, but for MS 2254.9-3712.}
\end{figure*}
\end{document}